\documentclass[%
 aip,
pof,
 amsmath,amssymb,
preprint,%
]{revtex4-1}

\usepackage {CJK}

\usepackage{graphicx}% Include figure files
\usepackage{dcolumn}% Align table columns on decimal point
\usepackage{bm}% bold math
\usepackage[utf8]{inputenc}
\usepackage[T1]{fontenc}
\usepackage{mathptmx}

\usepackage{url}

\begin{document}

\begin{CJK*}{UTF8}{} % Use default fonts from CJK (see below)

\preprint{Preprint submit to \emph{Physics of Fluids}}

\title{Statistics of temperature and thermal energy dissipation rate in low-Prandtl number turbulent thermal convection}
% Force line breaks with \\

\author{Ao Xu}
 \affiliation{School of Aeronautics, Northwestern Polytechnical University, Xi'an 710072, China}%

\author{Le Shi}
 \affiliation{State Key Laboratory of Electrical Insulation and Power Equipment, Center of Nanomaterials for Renewable Energy, School of Electrical Engineering, Xi'an Jiaotong University, Xi'an 710049, China}%

\author{Heng-Dong Xi}
 \email{hengdongxi@nwpu.edu.cn}
 \affiliation{School of Aeronautics, Northwestern Polytechnical University, Xi'an 710072, China}%

\date{\today}% It is always \today, today,

\begin{abstract}
We report the statistical properties of temperature and thermal energy dissipation rate in low-Prandtl number turbulent Rayleigh-B\'enard convection.
High resolution two-dimensional direct numerical simulations were carried out for the Rayleigh number ($Ra$) of $10^{6} \le Ra \le 10^{7}$ and the Prandtl number ($Pr$) of 0.025.
Our results show that the global heat transport and momentum scaling in terms of Nusselt number ($Nu$) and Reynolds number ($Re$) are  $Nu=0.21Ra^{0.25}$ and  $Re=6.11Ra^{0.50}$, respectively, indicating that the scaling exponents
are smaller than those for moderate-Prandtl number fluids (such as water or air) in the same convection cell.
In the central region of the cell, probability density functions (PDFs) of temperature profiles show stretched exponential peak and the Gaussian tail;
in the sidewall region, PDFs of temperature profiles show a multimodal distribution at relative lower $Ra$, while they approach the Gaussian profile at relative higher $Ra$.
We split the energy dissipation rate into contributions from bulk and boundary layers and found the \emph{locally averaged} thermal energy dissipation rate from the boundary layer region is an order of magnitude larger than that from the bulk region.
Even if the much smaller volume occupied by the boundary layer region is considered, the \emph{globally averaged} thermal energy dissipation rate from the boundary layer region is still larger than that from the bulk region.
We further numerically determined the scaling exponents of globally averaged thermal energy dissipation rates as functions of $Ra$ and $Re$.
\footnote{This article may be downloaded for personal use only.
Any other use requires prior permission of the author and AIP Publishing.
This article appeared in Xu et al., Phys. Fluids \textbf{31}, 125101 (2019) and may be found at \url{https://doi.org/10.1063/1.5129818}.}
\end{abstract}

\maketitle
\end{CJK*}

\section{\label{sec:level1}Introduction}

Thermal convection occurs ubiquitously in nature and has wide applications in industry.
When the kinematic viscosity of the fluid is much smaller than its thermal diffusivity, the thermal convective flow is called low-Prandtl number convection.
Low-Prandtl number convection has found its unique applications in the outer envelope of the Sun \cite{hanasoge2016seismic}, the liquid metal core of the Earth and other planets \cite{king2013turbulent}, the fission reactors of nuclear engineering \cite{grotzbach2013challenges}, the electrodes of liquid metal batteries \cite{kelley2014mixing}, and so on.
A paradigm for the study of thermal convection is the Rayleigh-B\'enard convection, which is a fluid layer heated from the bottom and cooled from the top \cite{ahlers2009heat,lohse2010small,chilla2012new,xia2013current,mazzino2017two}.
Challenges on laboratory experiments of low-Prandtl number convection mainly arise from opaque nature of the working fluid, which is usually liquid metal, excluding optical imaging techniques such as particle image velocimetry or Lagrangian particle tracking.
As for direct numerical simulations (DNS), extensive computational resources are needed to resolve the very thin viscous boundary layer in low-Prandtl number convection, where the production of vorticity and shear are strongly enhanced.
Due to the above reasons, previous studies on convection in low-Prandtl number fluids (such as liquid mercury and liquid gallium) are relative fewer compared with that in moderate-Prandtl number fluids (such as water and air).
Recent research progress on low-Prandtl number convection includes
Vogt et al.'s \cite{vogt2018jump} discovery that large-scale circulation takes the form of a jump rope vortex in cells of aspect ratio higher than unity when using liquid gallium as the working fluid.
Schumacher et al. \cite{schumacher2015enhanced} found that the generation of small-scale vorticity in the bulk convection follows the same mechanisms as idealized isotropic turbulence for low-Prandtl number convection.
Scheel and Schumacher \cite{king2013turbulent} identified a transition between the rotationally constrained and the weakly rotating turbulent states in rotating Rayleigh-B\'enard convection with liquid gallium that differs substantially from moderate-Prandtl number convection.
The main differences are due to the more diffuse temperature field, more vigorous velocity field, and coarser yet fewer production of thermal plumes in low-Prandtl number convection \cite{schumacher2015enhanced,scheel2016global,zwirner2018confined}.

In thermal convection, the global heat transport of the system is measured by the Nusselt number ($Nu$), which is defined as  $Nu=Q/(\chi\Delta_{T}/H)$.
Here, $Q$ is the heat current density across the fluid layer of height $H$,  $\chi$ is the thermal conductivity of the fluid, and $\Delta_{T}$ is the imposed temperature difference.
The control parameters of the system include the Rayleigh number ($Ra$), which describes the strength of buoyancy force relative to thermal and viscous dissipative effects via  $Ra=\beta g\Delta_{T}H^{3}/(\nu \kappa)$,
and the Prandtl number ($Pr$), which describes the thermophysical fluid properties via  $Pr=\nu / \kappa$.
Here, $\beta$, $\kappa$ and  $\nu$ are the thermal expansion coefficient, thermal diffusivity, and kinematic viscosity of the fluid, respectively.
$g$ is the gravitational acceleration.
In turbulent thermal convection, the energy supplied at large scales cascades to intermediate scales and then to dissipative scales.
To quantify the dissipation of kinetic and thermal energies due to fluid viscosity and thermal diffusivity, the kinetic and thermal energy dissipation rates are defined as $\varepsilon_{u}(\mathbf{x},t)=(\nu/2)\sum_{ij}[\partial_{i}u_{j}(\mathbf{x},t)+\partial_{j}u_{i}(\mathbf{x},t)]^{2}$  and  $\varepsilon_{T}(\mathbf{x},t)=\kappa\sum_{i}[\partial_{i}T(\mathbf{x},t)]^{2}$, respectively.
Shraiman and Siggia \cite{shraiman1990heat} averaged the equations of motion and derived exact relations of global averages of  $\varepsilon_{u}=\langle \varepsilon_{u}(\mathbf{x},t) \rangle_{V}$ and $\varepsilon_{T}=\langle \varepsilon_{T}(\mathbf{x},t) \rangle_{V}$.
The rigorous global exact relations of   $\varepsilon_{u}=\nu^{3}L^{-4}(Nu-1)RaPr^{-2}$ and  $\varepsilon_{T}=\kappa\Delta_{T}^{2}L^{-2}Nu$ further form the backbone of the Grossman-Lohse (GL) theory on turbulent heat transfer \cite{grossmann2000scaling,grossmann2002prandtl}.
In the GL theory, the energy dissipation rate was split into contributions from bulk and boundary layers, such that the scaling of $Nu$ and $Re$ in the $Ra$-$Pr$ phase diagram was obtained.
Later, Grossmann and Lohse  \cite{grossmann2004fluctuations} extended the GL theory and considered the role of thermal plumes.
They split $\varepsilon_{T}$ into contributions from turbulent background and plumes.
Although these two approaches to split energy dissipation involve different physical pictures about the local dynamics of turbulent convection, there is no change in the quantitative functional forms of $Nu$ and $Re$ with $Ra$ and $Pr$.
Based on the analysis of direct numerical simulation data, Emran and Schumacher \cite{emran2008fine} found, for $Pr = 0.7$ fluid, the probability density functions (PDFs) of  $\varepsilon_{T}$ in a cylindrical cell deviate from a log-normal distribution, but fit well by a stretched exponential distribution similar to passive scalar dissipation rate in homogeneous isotropic turbulence \cite{overholt1996direct}.
Kaczorowski and Wagner \cite{kaczorowski2009analysis} analyzed the contributions of bulk and boundary layers and plumes to the PDFs of the thermal dissipation rate in a rectangular cell of $Pr = 0.7$ fluid.
They found the core region scaling changes from pure exponential to a stretched exponential scaling with the increasing of $Ra$.
Zhang et al. \cite{zhang2017statisticsJFM} investigated statistical properties of   $\varepsilon_{u}$ and $\varepsilon_{T}$  in a two-dimensional square cell with $Pr = 0.7$ and $Pr = 5.3$ fluids.
They found the ensemble average of the scale of both dissipation rates as $Ra^{-0.18 \sim -0.20}$, in agreement with the prediction of global exact relations \cite{shraiman1990heat}.
The boundary layer and plume contributions scale as GL theory predictions, while the bulk and background contributions deviate from the GL theory predictions.
Within the viscous and thermal boundary layers, the PDFs of kinetic and thermal energy dissipation rates are non-log-normal and obey approximately a Bramwell-Holdsowrth-Pinton distribution \cite{zhang2017statisticsPRE}.
Bhattacharya et al. \cite{bhattacharya2018complexity} derived scaling relations for the viscous dissipation rate and viscous dissipation, and their results indicate that although the viscous dissipation rate  in the boundary layers is more intense, the viscous dissipation  in the bulk is larger than that in the boundary layers, which is caused by the large volume of the bulk region.

In this work, we quantify the statistics of the temperature and the thermal energy dissipation rate in low-Prandtl number Rayleigh-B\'enard convection, to further enrich our understandings of the flow dynamics and energy cascade in low-Prandtl number turbulent convection.
Here, we choose the working fluid with $Pr=0.025$ as an example, which corresponds to the typical Prandtl number of liquid gallium or mercury.
In contrast to conventional direct numerical simulation (DNS) based on solving the discretized nonlinear Navier-Stokes equations, we adopt the lattice Boltzmann (LB) method as an alternative numerical tool for DNS mainly due to two reasons.
One is that LB method is easy to be implemented and parallelized, benefiting from its local nonlinearity, while the other is that LB method has lower numerical dissipation compared to conventional second-order computational fluid dynamics methods \cite{chen1998lattice,aidun2010lattice,xu2017lattice}.
During the past several decades, the LB method has been successfully applied to DNS of turbulent flows, including decaying homogeneous isotropic turbulence \cite{yu2005dns,wang2016comparison}, turbulent channel and pipe flows \cite{peng2018direct,peng2019directJFM}, and turbulent thermal convective flows \cite{xu2019lattice}.
The rest of this paper is organized as follows:
In Sec. \ref{Section2}, we first present the mathematical model for the incompressible thermal flow under the Boussinesq approximation, followed by the LB method to obtain velocity and temperature fields.
In Sec. \ref{Section3}, we first introduce global features in low-Prandtl number thermal convection and then analyze the statistics of temperature and thermal energy dissipation rate.
In Sec. \ref{SectionConclusions}, main conclusions of the present work are summarized.

\section{\label{Section2}Numerical method}

\subsection{Mathematical model for incompressible thermal flow}

We consider incompressible thermal flows under the Boussinesq approximation.
The temperature is treated as an active scalar and its influence on the velocity field is realized through the buoyancy term.
The viscous heat dissipation and compression work are neglected, and all the transport coefficients are assumed to be constants.
The governing equations can be written as
\begin{subequations}
\begin{align}
& \nabla \cdot \mathbf{u}=0 \\
& \frac{\partial \mathbf{u}}{\partial t}+\mathbf{u}\cdot \nabla \mathbf{u}=-\frac{1}{\rho_{0}}\nabla p+\nu \nabla^{2}\mathbf{u}+g\beta(T-T_{0})\hat{\mathbf{y}} \\
& \frac{\partial T}{\partial t}+\mathbf{u}\cdot \nabla T=\kappa \nabla^{2} T
\end{align} \label{Eq.NS}
\end{subequations}
where $\mathbf{u}=(u,v)$ is the fluid velocity.
$p$ and $T$ are the pressure and temperature of the fluid, respectively.
$\rho_{0}$ and $T_{0}$  are the reference density and temperature, respectively.
$\hat{\mathbf{x}}$ and $\hat{\mathbf{y}}$  are the unit vectors in the horizontal and vertical directions, respectively.
With the scaling
\begin{equation}
    \begin{split}
&  \mathbf{x}/H \rightarrow \mathbf{x}^{*}, \ \ t/\sqrt{H/(g\beta\Delta_{T})} \rightarrow t^{*}, \ \ \mathbf{u}/\sqrt{g\beta H \Delta_{T}} \rightarrow \mathbf{u}^{*}, \\
& p/(\rho_{0}g \beta \Delta_{T}H) \rightarrow p^{*}, \ \ (T-T_{0})/\Delta_{T} \rightarrow T^{*}
    \end{split}
\end{equation}
then Eq. \ref{Eq.NS} can be rewritten in dimensionless form as
\begin{subequations}
\begin{align}
& \nabla \cdot \mathbf{u}^{*}=0 \\
& \frac{\partial \mathbf{u}^{*}}{\partial t}+\mathbf{u}^{*}\cdot \nabla \mathbf{u}^{*}=-\nabla p^{*}+\sqrt{\frac{Pr}{Ra}} \nabla^{2}\mathbf{u}^{*}+T^{*}\tilde{\mathbf{y}} \\
& \frac{\partial T^{*}}{\partial t}+\mathbf{u}^{*}\cdot \nabla T^{*}=\sqrt{\frac{1}{PrRa}} \nabla^{2} T
\end{align}
\end{subequations}

\subsection{The LB model for fluid flows and heat transfer}
The LB model to solve fluid flows and heat transfer is based on the double distribution function approach, which consists of a D2Q9 model for the Navier-Stokes equations (i.e., Eqs. 1a and 1b) to simulate fluid flows and a D2Q5 model for the convection-diffusion equations (i.e., Eq. 1c) to simulate heat transfer.
In the LB method, to solve Eqs. 1a and 1b, the evolution equation of the density distribution function is written as
\begin{equation}
  f_{i}(\mathbf{x}+\mathbf{e}_{i}\delta_{t},t+\delta_{t})-f_{i}(\mathbf{x},t)=-(\mathbf{M}^{-1}\mathbf{S})_{ij}\left[\mathbf{m}_{j}(\mathbf{x},t)-\mathbf{m}_{j}^{(\text{eq})}(\mathbf{x},t)\right]
  +\delta_{t}F_{i}^{'} \label{Eq.MRT}
\end{equation}
where $f_{i}$ is the density distribution function.
$\mathbf{x}$ is the fluid parcel position, $t$ is the time, and $\delta_{t}$ is the time step.
$\mathbf{e}_{i}$ is the discrete velocity along the $i$th direction.
$\mathbf{M}$ is a $9\times 9$ orthogonal transformation matrix that projects the density distribution function $f_{i}$ and its equilibrium $f_{i}^{(\text{eq})}$ from the velocity space onto the moment space, such that $\mathbf{m}=\mathbf{M}\mathbf{f}$ and $\mathbf{m}^{(\text{eq})}=\mathbf{M}\mathbf{f}^{(\text{eq})}$.
$\mathbf{S}=\text{diag}(s_{\rho},s_{e},s_{\varepsilon},s_{j},s_{q},s_{j},s_{q},s_{\nu},s_{\nu})$ is the diagonal relaxation matrix, where the relaxation parameters are chosen as
$s_{\rho}=s_{j}=0,\ s_{e}=s_{\varepsilon}=s_{\nu}=1/\tau_{f},\ s_{q}=8(2\tau_{f}-1)(8\tau_{f}-1)$. Here, $\tau_{f}$ is related with the kinematic viscosity of the fluid via $\nu=c_{s}^{2}(\tau_{f}-0.5)$.
The forcing term $F_{i}^{'}$ in the right-hand side of Eq. \ref{Eq.MRT} is given by $ \mathbf{F}^{'}=\mathbf{M}^{-1}\left( \mathbf{I}-\mathbf{S}/2 \right)\mathbf{M}\tilde{\mathbf{F}}$,
and the term $\mathbf{M\tilde{F}}$ is \cite{guo2002discrete}
\begin{equation}
\mathbf{M}\bar{\mathbf{F}}=\left[0, \ 6\mathbf{u}\cdot\mathbf{F}, \ -6\mathbf{u}\cdot\mathbf{F}, \ F_{x}, \ -F_{x}, \ F_{y}, \ -F_{y}, \ 2uF_{x}-2vF_{y}, \ uF_{x}+vF_{y} \right]^{T}
\end{equation}
where $\mathbf{F}=\rho g\beta(T-T_{0})\hat{\mathbf{y}}$.
The macroscopic density $\rho$ and velocity $\mathbf{u}$ are obtained from $\rho=\sum_{i=0}^{8}f_{i}, \ \ \mathbf{u}=\frac{1}{\rho}\left( \sum_{i=0}^{8}\mathbf{e}_{i}f_{i}+\mathbf{F}/2 \right)$.
To solve Eq. \ref{Eq.NS}c, the evolution equation of temperature distribution function is written as
\begin{equation}
  g_{i}(\mathbf{x}+\mathbf{e}_{i}\delta_{t},t+\delta_{t})-g_{i}(\mathbf{x},t)=-(\mathbf{N}^{-1}\mathbf{Q})_{ij}\left[\mathbf{n}_{j}(\mathbf{x},t)-\mathbf{n}_{j}^{(\text{eq})}(\mathbf{x},t)\right]
  \label{Eq.MRT_T}
\end{equation}
where $g_{i}$ is the temperature distribution function.
$\mathbf{N}$ is a $5\times 5$ orthogonal transformation matrix that projects the temperature distribution function $g_{i}$ and its equilibrium $g_{i}^{(\text{eq})}$ from the velocity space onto the moment space, such that $\mathbf{n}=\mathbf{N}\mathbf{g}$ and $\mathbf{n}^{(\text{eq})}=\mathbf{N}\mathbf{g}^{(\text{eq})}$.
$\mathbf{Q}=\text{diag}(0,q_{\kappa},q_{\kappa},q_{e},q_{\nu})$ is the diagonal relaxation matrix.
To achieve the isotropy of the fourth-order error term \cite{dubois2009towards}, the relationship for relaxation parameters in D2Q5 model leads to $q_{\kappa}=3-\sqrt{3}$, $q_{e}=q_{\nu}=4\sqrt{3}-6$ and  $a_{T}=20\sqrt{3}\kappa-6$, where  $a_{T}$ is a constant in the equilibrium distribution function $\mathbf{n}^{(\text{eq})}$.
The macroscopic temperature $T$ is obtained from $T=\sum_{i=0}^{4}g_{i}$.
More numerical details on the lattice Boltzmann method can be found in Ref.~\onlinecite{wang2013lattice,contrino2014lattice,xu2017accelerated,xu2019lattice}.

\subsection{\label{subsectionSettings}Simulation settings}

The top and bottom walls of the convection cell are kept at constant cold and hot temperatures, respectively; while the other two vertical walls are adiabatic.
All four walls impose no-slip velocity boundary condition.
The dimension of the cell is $L \times H$, and we set $L = H$ in this work.
Simulation results are provided for the Rayleigh number of $10^{6} \le Ra \le 10^{7}$ and the Prandtl number of $Pr = 0.025$.
To make sure that the statistically stationary state has been reached and the initial transient effects are washed out, the simulation protocol is as follows:
we first check whether statistically stationary state has reached in every $100 t_{f}$;
after that we check whether statistically convergent state has reached in every $100 t_{f}$.
The averaging time $t_{avg}$ to obtain statistically convergent results are given in Table \ref{Table1}.
Here, $t_{f}$  denotes the free-fall time unit  $t_{f}=\sqrt{H/(g\beta\Delta_{T})}$.
We also check whether the grid spacing $\Delta_{g}$ and time interval $\Delta_{t}$ is properly resolved by comparing with the Kolmogorov and Batchelor scales.
The Kolmogorov length scale is estimated by the global criterion  $\eta=HPr^{1/2}/[Ra(Nu-1)]^{1/4}$, the Batchelor length scale is estimated by  $\eta_{B}=\eta Pr^{-1/2}$, and the Kolmogorov time scale is estimated as  $\tau_{\eta}=\sqrt{\nu/\langle \varepsilon_{u} \rangle}=\sqrt{Pr/(Nu-1)}$.
The global heat transport is measured as the volume averaged Nusselt number as  $Nu_{vol}=1+\sqrt{Pr Ra}\langle vT \rangle_{V,t}$.
From Table \ref{Table1}, we can see that grid spacing satisfy  $\max(\Delta_{g}/\eta, \Delta_{g}/\eta_{B}) \le 0.45$, which ensures the spatial resolution.
In addition, the time intervals are  $\Delta_{t} \le 0.00055 \tau_{\eta}$, thus guaranteeing an adequate temporal resolution.

\begin{table}
\caption{Spatial and temporal resolutions of the simulations.}
\begin{ruledtabular}
\begin{tabular}{ccccccc}
$Ra$	& $Pr$	& Mesh size	& $\Delta_{g}/\eta$	& $\Delta_{g}/\eta_{B}$	& $\Delta_{t}/\tau_{\eta}$	& $t_{avg}/t_{f}$ \\
\hline
$1.0\times10^{6}$	& 0.025	& $769^{2}$	    & 0.39	& 0.062	& $5.45\times10^{-4}$	& 700  \\
$1.3\times10^{6}$	& 0.025	& $851^{2}$	    & 0.39	& 0.062	& $5.22\times10^{-4}$	& 2300 \\
$1.6\times10^{6}$	& 0.025	& $901^{2}$	    & 0.40	& 0.063	& $5.11\times10^{-4}$	& 1400 \\
$2.0\times10^{6}$	& 0.025	& $1001^{2}$	& 0.38	& 0.060	& $4.73\times10^{-4}$	& 1000 \\
$3.0\times10^{6}$	& 0.025	& $1025^{2}$	& 0.43	& 0.067	& $4.89\times10^{-4}$	& 1400 \\
$4.0\times10^{6}$	& 0.025	& $1101^{2}$	& 0.43	& 0.068	& $4.71\times10^{-4}$	& 1000 \\
$6.0\times10^{6}$	& 0.025	& $1201^{2}$	& 0.45	& 0.071	& $4.54\times10^{-4}$	& 1300 \\
%$8.0\times10^{6}$	& 0.025	& $1355^{2}$	& 0.44	& 0.069	& $4.18\times10^{-4}$	& 1000 \\
$1.0\times10^{7}$	& 0.025	& $1537^{2}$	& 0.42	& 0.066	& $3.83\times10^{-4}$	& 1000 \\
\end{tabular}
\end{ruledtabular} \label{Table1}
\end{table}

In Rayleigh-B\'enard convection, in addition to the volume averaged Nusselt number, we can define the average Nusselt number over top and bottom walls as  $Nu_{wall}=-1/2(\langle \partial_{z}T \rangle_{top,t}+\langle \partial_{z}T \rangle_{bottom,t})$ and the thermal energy dissipation rate based Nusselt number as  $Nu_{th}=\sqrt{Ra Pr}\langle \varepsilon_{T} \rangle_{V,t}$.
If the direct numerical simulation of RB convection is well resolved and statically convergent, the above three definitions of Nusselt numbers should give the same result.
Here, the volume averaged Nusselt number $Nu_{vol}$ is chosen as the reference value to calculate its relative differences with other Nusselt numbers, and the differences (denoted by 'diff.') are included in brackets in the corresponding columns.
From Table \ref{Table2}, we can see the differences are around 1\%, indicating that Nusselt numbers show good consistency with each other.
In addition, to measure global strength of the convection, the Reynolds number based on root-mean-square (rms) velocity is defined as $Re=\sqrt{\langle u^{2}+v^{2} \rangle_{V,t}}H/\nu$.
Even for convective turbulence at the moderate Rayleigh number of $Ra=10^{7}$, the corresponding Reynolds number of the turbulent flow can reach  $Re \sim O(10^{4})$.

\begin{table}
\caption{Nusselt and Reynolds numbers as a function of Rayleigh number.}
\begin{ruledtabular}
\begin{tabular}{ccccccc}
$Ra$	& $Pr$	& $Nu_{vol}$	& $Nu_{wall}$ (diff.)	& $Nu_{th}$ (diff.)	& $Re$ \\
\hline
$1.0\times10^{6}$	& 0.025	& 6.28	& 6.25 (0.38\%)	& 6.35 (1.12\%)	& 6025.06 \\
$1.3\times10^{6}$	& 0.025	& 6.92	& 6.97 (0.79\%)	& 7.01 (1.27\%)	& 7064.86 \\
$1.6\times10^{6}$	& 0.025	& 7.37	& 7.36 (0.12\%)	& 7.32 (0.59\%)	& 7789.17 \\
$2.0\times10^{6}$	& 0.025	& 7.72	& 7.74 (0.28\%)	& 7.73 (0.12\%)	& 8779.49 \\
$3.0\times10^{6}$	& 0.025	& 8.54	& 8.57 (0.37\%)	& 8.57 (0.43\%)	& 10628.02 \\
$4.0\times10^{6}$	& 0.025	& 9.08	& 9.13 (0.60\%)	& 9.14 (0.74\%)	& 12208.34 \\
$6.0\times10^{6}$	& 0.025	& 9.93	& 9.99 (0.60\%)	& 10.00 (0.70\%)	& 14843.71 \\
%$8.0\times10^{6}$	& 0.025	& 10.63	& 10.77 (1.36\%)	& 10.82 (1.76\%)	& 17192.23 \\
$1.0\times10^{7}$	& 0.025	& 11.38	& 11.39 (0.11\%)	& 11.42 (0.36\%)	& 19512.48 \\
\end{tabular}
\end{ruledtabular} \label{Table2}
\end{table}

\section{\label{Section3}Results and discussion}

\subsection{\label{Section31}Global features}

A typical snapshot of an instantaneous flow field and the corresponding temperature, vorticity and logarithmic thermal energy dissipation rate fields are shown in Fig. \ref{Fig1},
and a corresponding video can be viewed in the supplementary material.
At the same $Ra$, low-Prandtl number turbulent thermal convection is more vigorous due to inertial effects.
The temperature field is diffusive with the coarse plumes near the top and bottom boundary layers,
rather than the filamented plumes in moderate-Prandlt number turbulent thermal convection.
The production of vorticity is strong near all the four walls, while intense dissipations of thermal energy occur in regions of detached hot or cold plumes from bottom and top boundary layers, in consistent with previous studies  \cite{kerr1996rayleigh,shishkina2007local,emran2008fine,zhang2017statisticsJFM} that rising and falling thermal plumes are associated with large amplitudes of thermal energy dissipation rates.

\begin{figure}
\centering
\includegraphics[width=14cm]{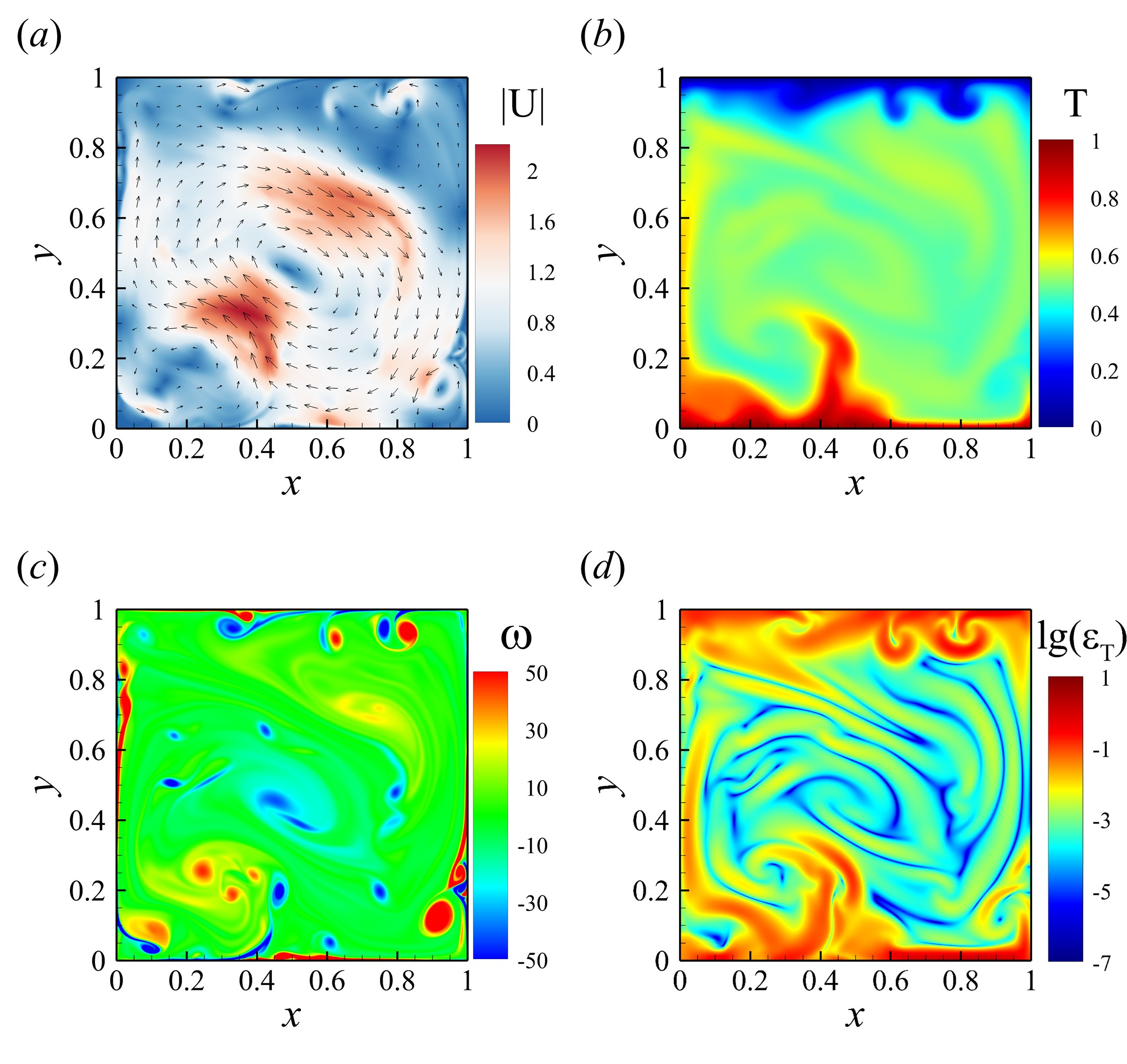}
\caption{\label{Fig1} A typical snapshot of an instantaneous flow field (a) and its corresponding temperature field (b), vorticity field (c), and logarithmic thermal energy dissipation rate field (d) for $Ra = 10^{7}$ and $Pr = 0.025$.}
\end{figure}

\begin{figure}
\centering
\includegraphics[width=14cm]{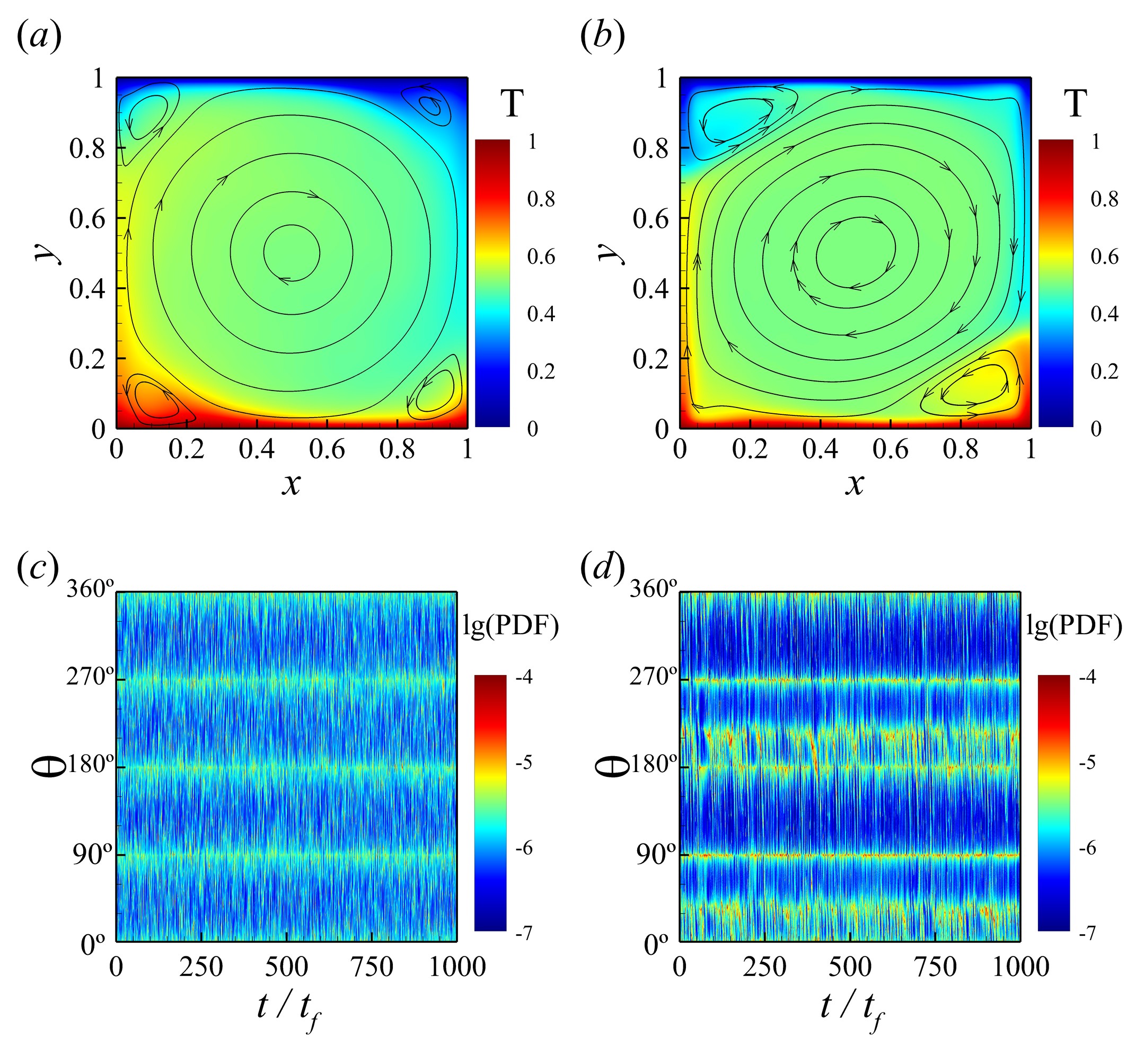}
\caption{\label{Fig2}  (a, b) Time-averaged temperature fields and streamlines, (c, d) time evolution (from left to right) of the probability density functions (PDFs) of the instantaneous velocity vector orientation $\theta$ for  $Pr = 0.025$ (a, c) and  $Pr = 5.3$ (b, d) at $Ra = 10^{7}$.}
\end{figure}

Time-averaged temperature fields and streamlines obtained at $Ra = 10^{7}$ for both low- and moderate Prandtl number convections are shown in Figs. \ref{Fig2}(a) and \ref{Fig2}(b), respectively.
Numerical details on DNS of moderate-Prandtl number convection can be found in the Appendix.
From Figs. \ref{Fig2}(a) and \ref{Fig2}(b), we can observe a typical flow pattern of Rayleigh-B\'enard convection, where there exists a well-defined LSC, together with counter-rotating corner rolls.
Meanwhile, we notice distinguishable differences on the flow pattern in this time-averaged flow field.
At low Prandtl number (i.e., $Pr = 0.025$), the LSC is in the form of a circle, and there exist four secondary corner vortices;
at moderate Prandtl number (i.e., $Pr = 5.3$), the LSC is in the form of a tilted ellipse, sitting along a diagonal of the flow cell with two secondary corner vortices that exist along the other diagonal.
A similar pattern was reported in a quasi-two-dimensional RB cell at a moderate Prandtl number. \cite{zhou2018similarity}
We further calculate the probability density functions (PDFs) of velocity vector orientation $\theta$ and plot its time evolution in Figs. \ref{Fig2}(c) and \ref{Fig2}(d).
Each vertical slice is a PDF of $\theta$ for an instantaneous velocity field.
Here, we count velocity of fluid nodes that belong to the inscribed circle region of the square convection cell.
Figure \ref{Fig2}(c) indicates the velocity vector orientations have  high probability values around $0^{\circ}$ (or $360^{\circ}$),  $90^{\circ}$, $180^{\circ}$, and $270^{\circ}$, implying the velocities of rising and falling thermal plumes, as well as horizontal 'wind'.
Comparing  with Figs. \ref{Fig2}(c) and \ref{Fig2}(d), we can find the velocity vector orientations have additional high probability values around $30^{\circ}$ and $210^{\circ}$, suggesting the diagonal orientation of the main roll for $Pr = 5.3$.

\begin{figure}
\centering
\includegraphics[width=14cm]{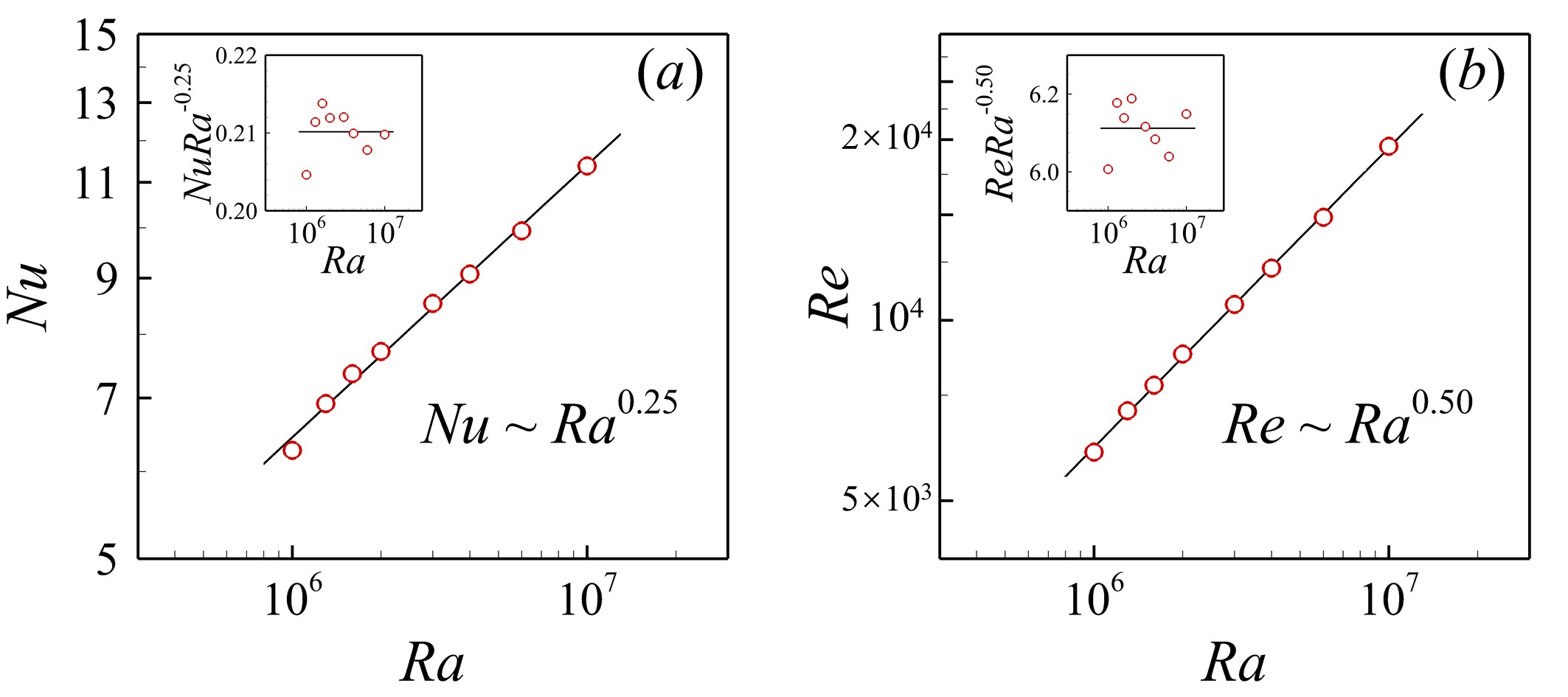}
\caption{\label{Fig3}   (a) Nusselt number and (b) Reynolds number as functions of Rayleigh number for $Pr = 0.025$. The solid lines are the power-law fits to the data. The insets are the compensated plots of the data shown in (a) and (b), respectively.}
\end{figure}

The measured Nusselt and Reynolds numbers as functions of Rayleigh number are shown in Figs. \ref{Fig3}(a) and \ref{Fig3}(b), respectively.
The data can be well described by a power-law relation $Nu=0.21Ra^{0.25}$  and  $Re=6.11Ra^{0.50}$, indicated by the solid lines in the figures.
The heat transfer scaling exponent is in general consistent with previous experimental results and direct numerical simulation results obtained in a cylindrical RB cell filled with liquid mercury or liquid gallium   \cite{cioni1997strongly,king2013turbulent,scheel2016global,scheel2017predicting}, where $Nu \propto Ra^{0.25 \sim 0.27}$,
while the momentum scaling exponent from the present two-dimensional simulation is larger than that in previous three-dimensional simulations \cite{scheel2016global,scheel2017predicting}, where $Re \propto Ra^{0.44 \sim 0.45}$.
The above findings indicate that the $Nu(Ra)$ scaling exponent for two- and three-dimensional convection is very close, while the $Re(Ra)$ scaling exponent is larger in two-dimensional convection.
This trend is similar with previous comparison between two-and three-dimensional convections at moderate Prandtl number \cite{van2013comparison}.
The scaling exponents of $Nu(Ra)$ and $Re(Ra)$ for our low $Pr$ case are lower than those obtained in moderate $Pr$ cases, such as $Pr = 0.7$ case \cite{zhang2017statisticsJFM,zhang2018surface}, $Pr = 1.0$ case \cite{johnston2009comparison,van2012flow}, $Pr = 4.4$ case \cite{sugiyama2009flow} and $Pr = 5.3$ case \cite{zhang2017statisticsJFM}, where $Nu \propto Ra^{0.285 \sim 0.30}$   and $Re \propto Ra^{0.59 \sim 0.62}$,
while the prefactors of $Nu(Ra)$ and $Re(Ra)$ for the low $Pr$ case are larger than those obtained in the moderate $Pr$ case \cite{huang2013counter}.

\subsection{Statistics of temperature}

\begin{figure}
\centering
\includegraphics[width=14cm]{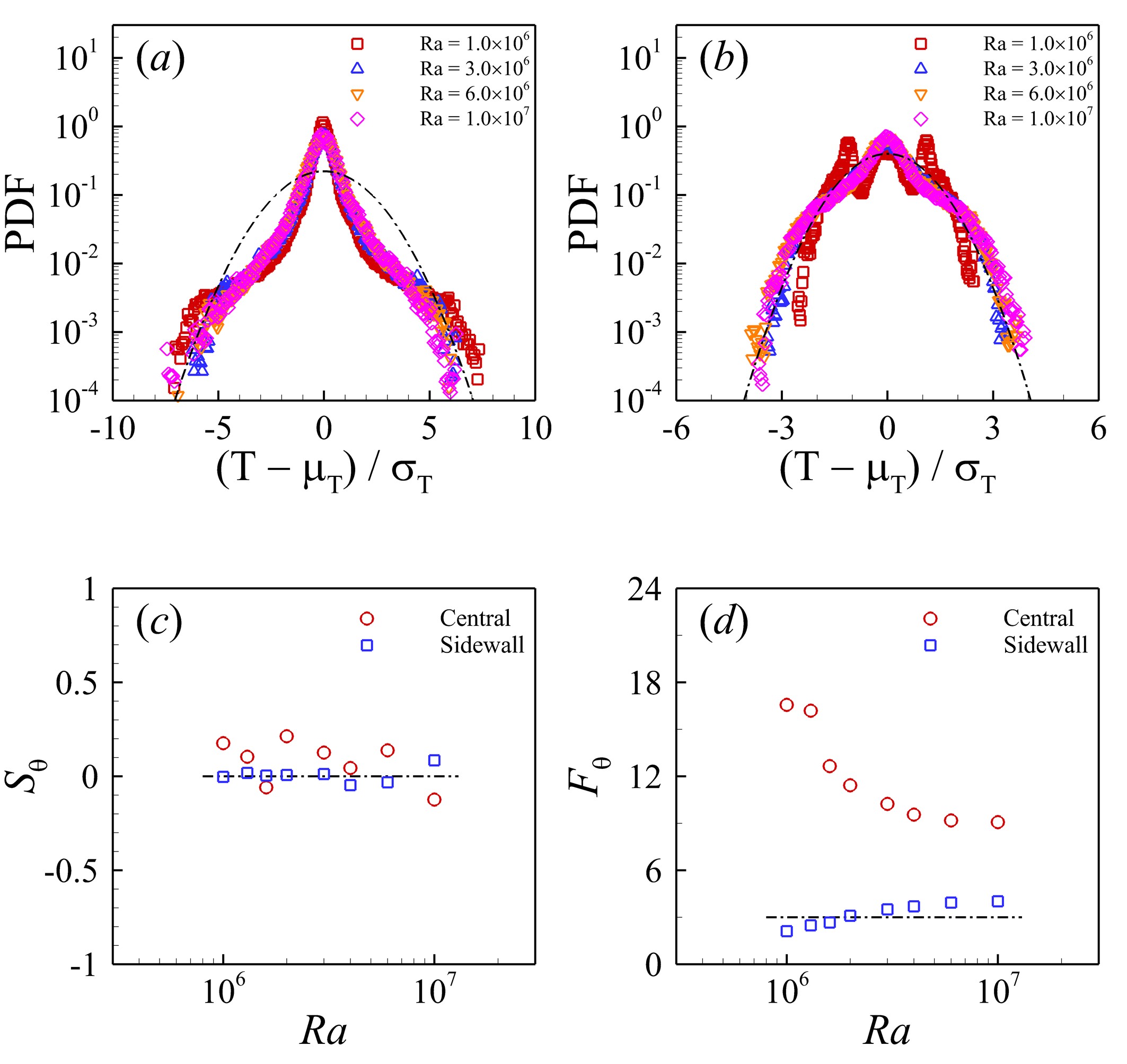}
\caption{\label{Fig4}
(a) Probability density functions (PDFs) of the normalized temperature $(T-\mu_{T})/\sigma_{T}$ measured along the line in central region at mid-height, i.e., $0.25 L \le x \le 0.75 L$ and $y = 0.5 H$, the dotted-dash line represents a Gaussian distribution;
(b) PDFs of the normalized temperature measured along the line in sidewall region at mid-height, i.e., $0 \le x \le 0.25 L$, $0.75 L \le x \le L$ and $y = 0.5H$;
(c) skewness of the temperature along the same line as that in (a) and (b), the dotted-dash line represents the value of zero;
(d) flatness of the temperature along the same line as that in (a) and (b), the dotted-dash line represents flatness for a Gaussian distribution.}
\end{figure}

We show the probability density functions (PDFs) of normalized temperature $(T-\mu_{T})/\sigma_{T}$  measured at the mid-height (i.e., $y = 0.5 H$) in two regions:
one is in the central region, i.e., $0.25 L \le x \le 0.75 L$, see Fig. \ref{Fig4}(a); the other is in the sidewall region, i.e., $0 \le x \le 0.25 L$ and $0.75 L \le x \le L$, see Fig. \ref{Fig4}(b).
Here, $\mu_{T}$ and $\sigma_{T}$ represent the mean value and standard deviation of $T$.
Generally, the temperature PDFs are symmetric at the mid-height of the convection cell, in agreement with previous findings at moderate-Prandtl number convection\cite{kerr1996rayleigh,emran2008fine}.
To quantitatively describe the asymmetry of the PDFs of the temperature, we calculate the skewness of temperature  $S_{\theta}$ as
\begin{equation}
S_{\theta} (y = 0.5H) =\frac{\langle \theta^{3} \rangle_{\mathbf{x},  t}}{ \langle \theta^{2} \rangle ^{3/2}_{\mathbf{x}, t}}
\end{equation}
where $\theta=T-\mu_{T}$.
The average $\langle \cdot \rangle_{\mathbf{x},t}$ is calculated over time $t$ and along the horizontal line $\mathbf{x}$ in the central or sidewall region.
From Fig. \ref{Fig4}(c), we can see that the skewness values are around zero in both central and sidewall regions,
indicating the rising hot plumes are comparable with falling cold plumes at the cell mid-height.
As for the shapes of the temperature PDFs profiles in the central region, the peaks show stretched exponential behavior and the tails show Gaussian behavior for all the considered Rayleigh numbers (indicated by the black dotted-dashed line, see Fig. \ref{Fig4}(a)).
In the sidewall region, the temperature PDF profiles show a multimodal distribution at relative lower Rayleigh number (e.g., $Ra =10^{6}$), indicating the flow state is in the regime transition to hard turbulence \cite{heslot1987transitions,castaing1989scaling};
at a relative higher Rayleigh number, the temperature PDF profiles approach the Gaussian profile indicated by the black dotted-dash line, see Fig. \ref{Fig4}(b).
To quantitatively describe the magnitude of the deviation from Gaussianity, we calculate the flatness of temperature  $F_{\theta}$ as
\begin{equation}
F_{\theta} (y = 0.5H) =\frac{\langle \theta^{4} \rangle_{\mathbf{x},  t}}{ \langle \theta^{2} \rangle ^{2}_{\mathbf{x}, t}}
\end{equation}
We can see from Fig. 4(d) that the flatness in central and sidewall regions show different trends with the increasing of Rayleigh number.
The large differences in these two regions are mainly due to the disparity in the number of plumes, since the central region has relative few plumes and the sidewall region is dominated by thermal plumes\cite{xi2004laminar}.

%%in the central region is much larger than 3 (i.e., the flatness value for a Gaussian distribution),

\subsection{Statistics of thermal energy dissipation rate}

\begin{figure}
\centering
\includegraphics[width=14cm]{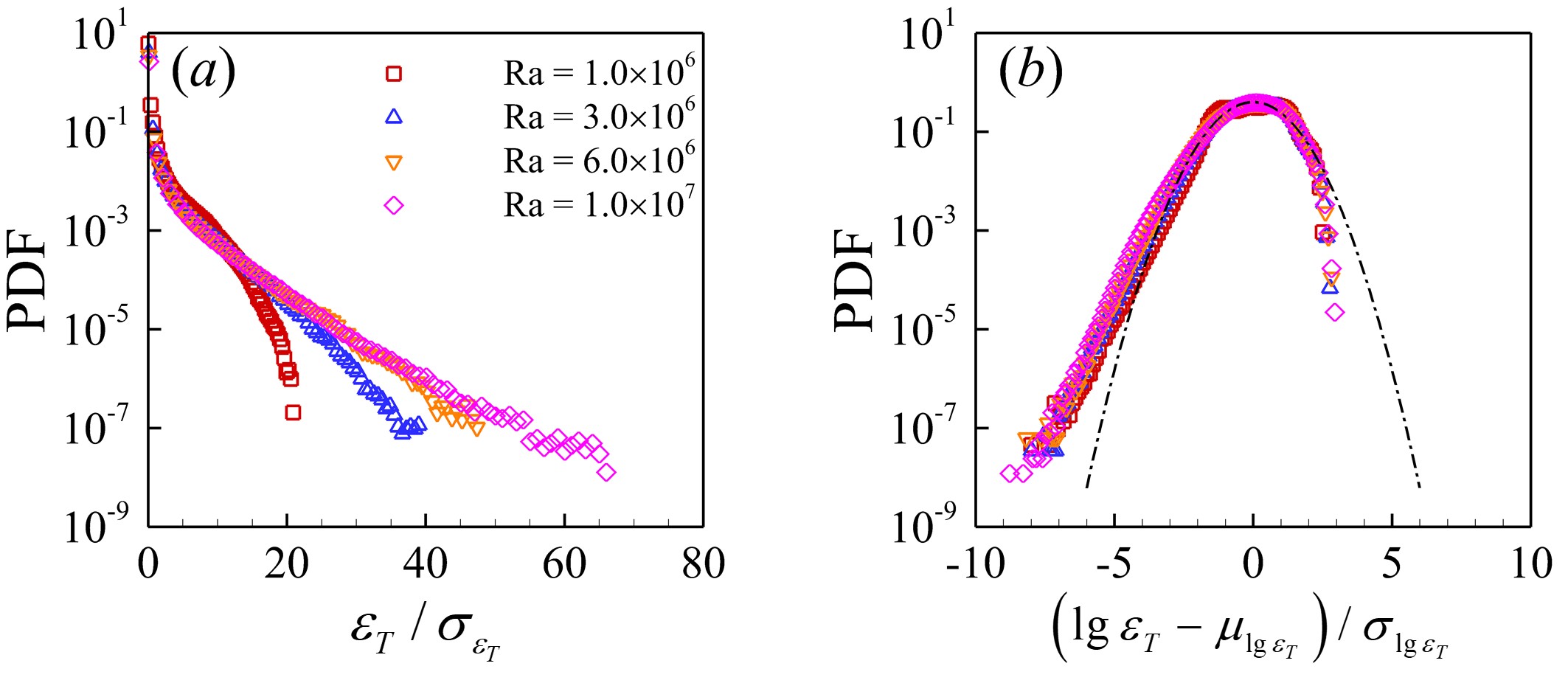}
\caption{\label{Fig5}   (a) Probability density functions (PDFs) of the thermal energy dissipation rate $\varepsilon_{T}(\mathbf{x},t)$, and (b) PDFs of the normalized logarithmic thermal energy dissipation rate $\lg \varepsilon_{T}(\mathbf{x},t)$  obtained over the whole cell, the dotted-dashed line represents a log-normal distribution.}
\end{figure}

Figure \ref{Fig5}(a) shows the PDFs of thermal energy dissipation rates $\varepsilon_{T}(\mathbf{x},t)$  obtained over the whole cell and over time, further normalized by their root-mean-square (rms) values.
The PDF tails become more extended with increasing of $Ra$, implying an increasing degree of small-scale intermittency of the thermal energy dissipation field.
We further check whether the thermal energy dissipation fields have a log-normal distribution as proposed by Kolmogorov \cite{kolmogorov1962refinement}.
Figure \ref{Fig5}(b) shows the PDFs of normalized logarithmic thermal energy dissipation rate $(\lg \varepsilon_{T}-\mu_{\lg \varepsilon_{T}})/\sigma_{\lg \varepsilon_{T}}$, and we can observe clear departures from log-normality for the thermal energy dissipation field, which is mainly due to the intermittent nature of local dissipation.
Similar observations have also been made for moderate-Prandtl number convection \cite{emran2008fine,zhang2017statisticsJFM,zhou2016kinetic} .

\begin{figure}
\centering
\includegraphics[width=14cm]{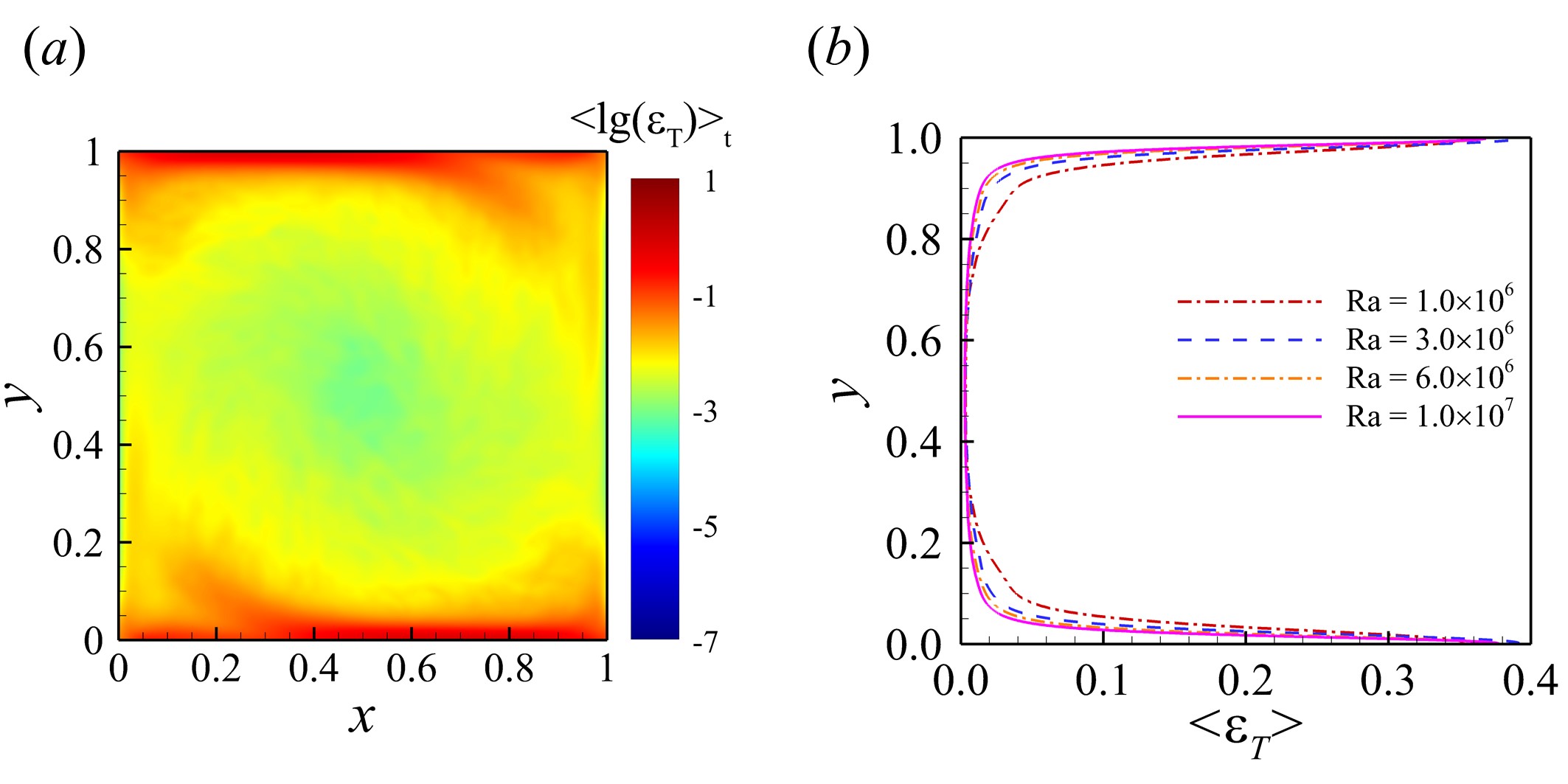}
\caption{\label{Fig6}   (a) Time-averaged logarithmic thermal energy dissipation field obtained at $Ra = 10^{7}$ and $Pr = 0.025$, and (b) vertical profiles of  horizontal- and time-averaged thermal energy dissipation rates for $Pr = 0.025$ and various $Ra$.}
\end{figure}

The time-averaged logarithmic thermal energy dissipation field $\langle \lg \varepsilon_{T}(\mathbf{x},t) \rangle$  obtained at $Ra = 10^{7}$ and $Pr = 0.025$ is shown in Fig. \ref{Fig6}(a).
From the time averaged field, it is seen that the contribution of thermal plumes to thermal energy dissipation is filtered out, and we can only see intense thermal energy dissipation occurs near the top and bottom walls where there are strong temperature gradients.
At the sidewall, the thermal energy dissipation rates do not increase significantly due to the adiabatic sidewall boundary conditions.
The vertical profiles of $\langle \varepsilon_{T}(\mathbf{x},t) \rangle_{x,t}(y)$  that averaged over the horizontal direction and over time are shown in Fig. \ref{Fig6}(b), which further illustrates the spatial distribution of thermal energy dissipation rate.
The thermal energy dissipation rate remains nearly zero in the bulk and increases rapidly near the top and bottom boundary layers, suggesting intense thermal energy dissipation within thermal boundary layers.

\begin{figure}
\centering
\includegraphics[width=14cm]{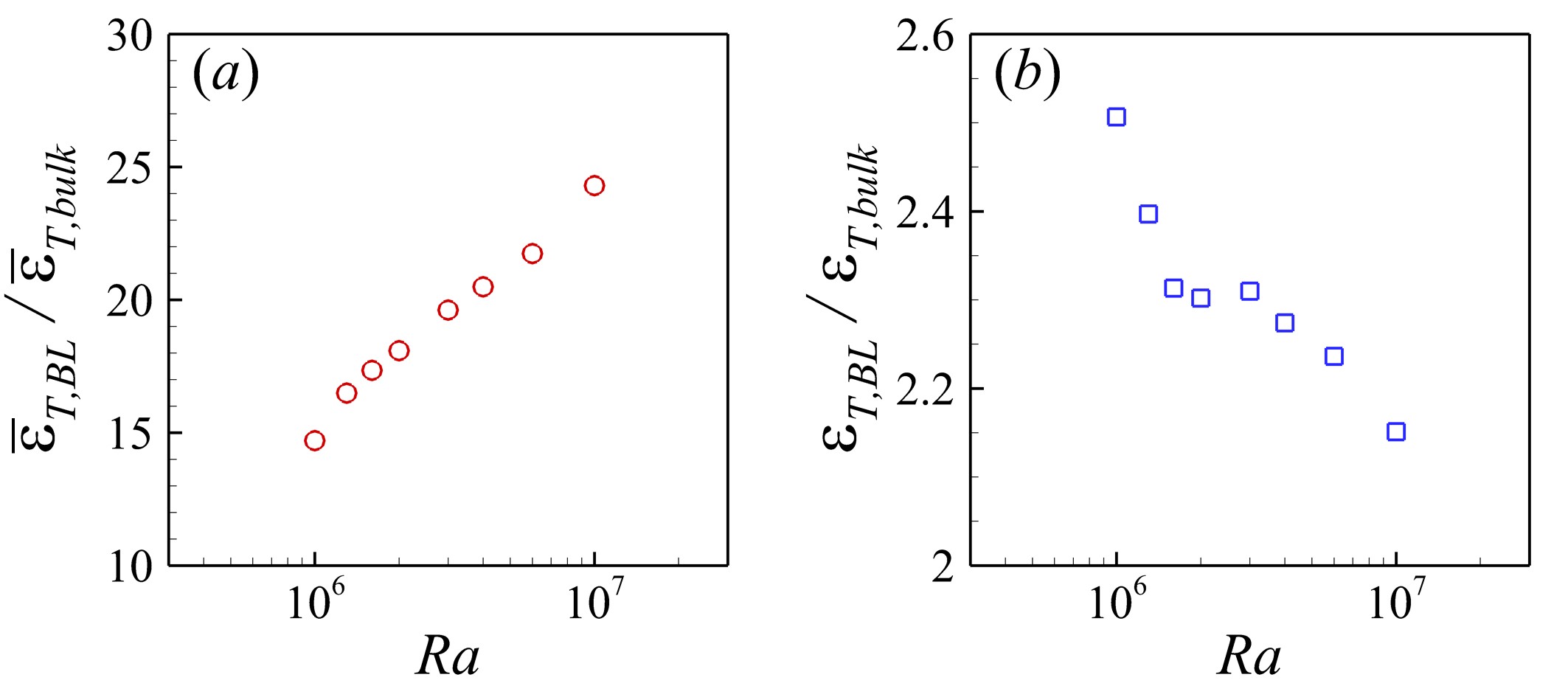}
\caption{\label{Fig7}   Ratio of (a) locally averaged and (b) globally averaged thermal energy dissipation rates from the thermal boundary layer and the bulk.}
\end{figure}

To quantitatively describe the spatial distribution of thermal energy dissipation, the thermal energy dissipation rate is partitioned into contributions from bulk and boundary layers, which is the essence of the Grossmann-Lohse (GL) theory on turbulent heat transfer \cite{grossmann2000scaling,grossmann2002prandtl}.
We first calculate the locally averaged thermal energy dissipation rates from the thermal boundary layer and the bulk as  $\bar{\varepsilon}_{T,BL}=\left[ \int_{0\le y \le \delta_{T}}+\int_{H-\delta_{T} \le y \le H}\kappa (\partial_{i}T)^{2}dy \right]/(2\delta_{T})=\kappa \langle (\partial_{i}T(\mathbf{x}\in BL,t))^{2} \rangle_{V_{BL}}$ and  $\bar{\varepsilon}_{T,bulk}=\left[ \int_{\delta_{T} \le y \le H-\delta_{T}} \kappa (\partial_{i}T)^{2}dy \right]/(H-2\delta_{T})=\kappa \langle (\partial_{i}T(\mathbf{x}\in bulk,t))^{2} \rangle_{V_{bulk}}$, respectively.
Figure \ref{Fig7}(a) shows the ratio of  $\bar{\varepsilon}_{T,BL}$ and $\bar{\varepsilon}_{T,bulk}$  as a function of Rayleigh number.
Here, the thermal boundary layer thickness  $\delta_{T}$ is determined as the distance between the wall and the position at which the rms temperature is maximum.
We can observe thermal energy dissipation rate that comes from the boundary layer region is an order of magnitude larger than that from the bulk region. With increasing Rayleigh number, thermal energy dissipation rate in the boundary layer is more intense.
We further calculate the globally averaged thermal energy dissipation rates from the thermal boundary layer and the bulk as  $\varepsilon_{T,BL}=\left[ \int_{0\le y \le \delta_{T}}+\int_{H-\delta_{T} \le y \le H}\kappa (\partial_{i}T)^{2}dy \right]/H=\kappa \langle (\partial_{i}T(\mathbf{x}\in BL,t))^{2} \rangle_{V}$ and  $\varepsilon_{T,bulk}=\left[ \int_{\delta_{T} \le y \le H-\delta_{T}} \kappa (\partial_{i}T)^{2}dy \right]/H=\kappa \langle (\partial_{i}T(\mathbf{x}\in bulk,t))^{2} \rangle_{V}$, respectively.
Figure \ref{Fig7}(b) shows the ratio of $\varepsilon_{T,BL}$  and $\varepsilon_{T,bulk}$  as a function of Rayleigh number.
Although the boundary layer region occupies much smaller volume than the bulk region, we can still observe that more thermal energy is dissipated in the boundary layer region compared to that in bulk region.

\begin{figure}
\centering
\includegraphics[width=14cm]{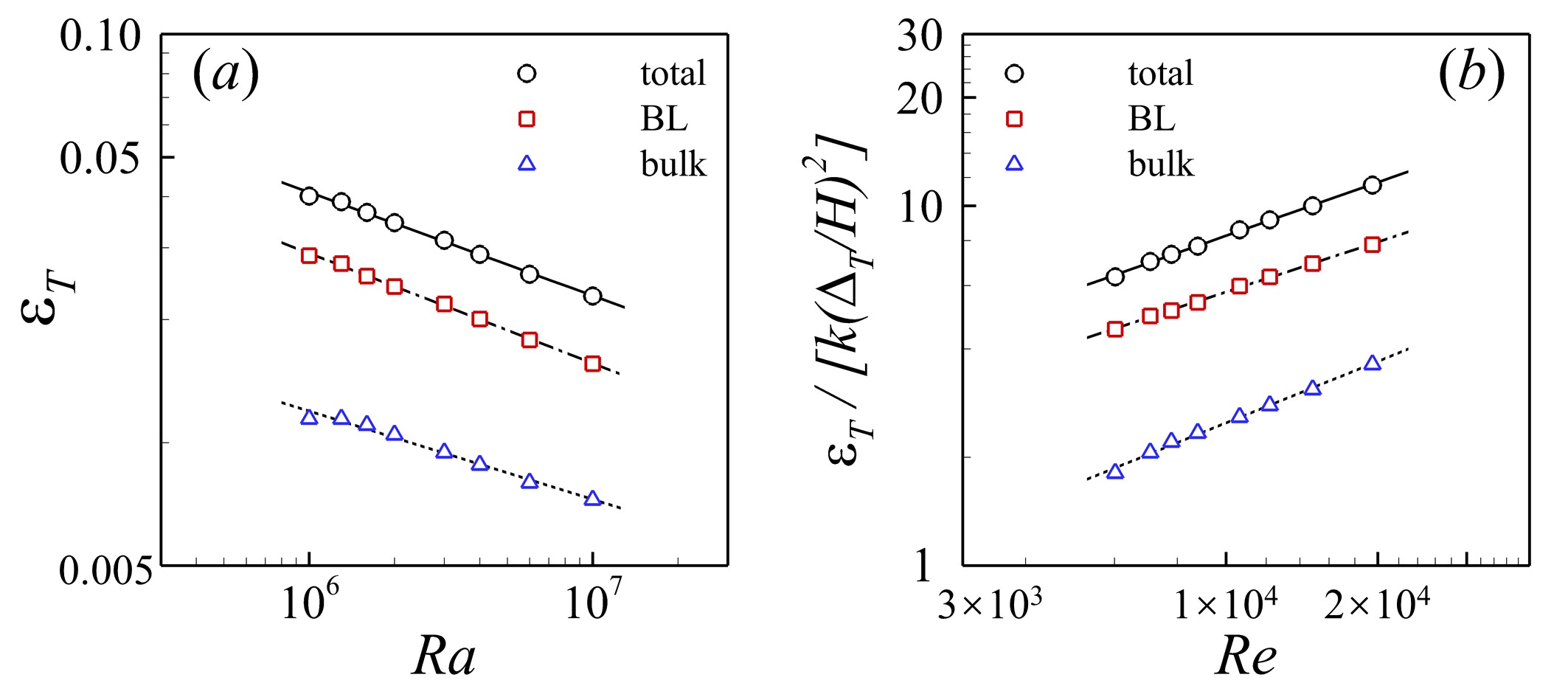}
\caption{\label{Fig8}   (a) Thermal energy dissipation rates as a function of $Ra$; (b) normalized thermal energy dissipation rates as a function of $Re$. The lines are the power-law fits to the corresponding data.}
\end{figure}

Globally averaged thermal energy dissipation rates as a function of Rayleigh number are shown in Fig. \ref{Fig8}(a).
For the total thermal energy dissipation rate over the whole cell, the data can be well described by a power-law relation  $\varepsilon_{T,total}=1.35Ra^{-0.25}$, indicated by the solid line in the figure.
This scaling exponent is larger than that for $Pr = 0.7$ and $Pr = 0.53$ obtained from direct numerical simulations in a two-dimensional cell \cite{zhang2017statisticsJFM}, where the exponent is -0.20.
On the other hand, the scaling behavior can be understood based on the global exact relation \cite{shraiman1990heat} of $\varepsilon_{T,total}=Nu/\sqrt{RaPr}$.
Since we have obtained  $Nu \sim Ra^{0.25}$ for $Pr = 0.025$ in Sec. \ref{Section31}, substitute the $Nu \sim Ra$ scaling into the global exact relation, we have  $\varepsilon_{T,total} \sim Ra^{-0.25}$.
The excellent agreement in the scaling exponent also demonstrates that the global exact relations are satisfied in our simulations.
For the thermal energy dissipation rates from the boundary layer and bulk, the scaling behavior can be described by  $\varepsilon_{T,BL}=1.20Ra^{-0.27}$ and $\varepsilon_{T,bulk}=0.24Ra^{-0.22}$, respectively.
Figure \ref{Fig8}(b) further shows the normalized globally averaged thermal dissipation rates  $\varepsilon_{T}/[\kappa(\Delta_{T}/H)^{2}]$ as a function of Reynolds number.
For the normalized total thermal energy dissipation rate over the whole cell, the data can be well described by a power-law relation  $\varepsilon_{T,total}/[\kappa (\Delta_{T}/H)^{2}] \sim Re^{0.49}$.
This scaling behavior can also be understood based on the global exact relation \cite{shraiman1990heat} of $\varepsilon_{T,total}=\kappa \Delta_{T}^{2}/H^{2}Nu$  as follows:
since we have obtained $Nu \sim Ra^{0.25}$  and $Re \sim Ra^{0.50}$   for $Pr = 0.025$ in Sec. \ref{Section31}, substitute the $Nu \sim Re^{0.50}$ relation into the global exact relation, we have  $\varepsilon_{T,total}/[\kappa(\Delta_{T}/H)^{2}] \sim Re^{0.50}$.
Again, the excellent agreement in the scaling exponent demonstrates that the global exact relations are satisfied in our simulations.
As for the boundary layer and bulk regions, compared with moderate-Prandtl number convection in the same convection cell \cite{zhang2017statisticsJFM}, in the current low-Prandtl case the scaling exponent of $\varepsilon_{T,BL}/[\kappa(\Delta_{T}/H)^{2}] \sim Re^{0.46}$  in the boundary layer region is slight smaller, while the scaling exponent of  $\varepsilon_{T,bulk}/[\kappa(\Delta_{T}/H)^{2}] \sim Re^{0.57}$ in the bulk region is significantly larger.

\section{\label{SectionConclusions}Conclusions}

In this work, we have presented high-resolution direct numerical simulations of a low-Prandtl number thermal convective flow and analyzed the statistical properties of temperature and thermal energy dissipation rate. The main findings are summarized as follows:

\begin{enumerate}
  \item
For low Prandtl number of $Pr = 0.025$, the global heat transport and momentum scaling are  $Nu=0.21Ra^{0.25}$ and $Re=6.11Ra^{0.50}$, respectively.
Both the exponents of $Nu(Ra)$ and $Re(Ra)$ are smaller than those for a moderate Prandtl number in the same convection cell.

  \item
Locally averaged thermal energy dissipation rate from the boundary layer region is an order of magnitude larger than that from the bulk region.
Even if the much smaller volume occupied by the boundary layer region is considered, the \emph{globally averaged} thermal energy dissipation rate from the boundary layer region is still larger than that from the bulk region.

  \item
The scaling exponents of globally averaged thermal energy dissipation rates with Rayleigh and Reynolds numbers are numerically determined as $\varepsilon_{T,total} \sim Ra^{-0.25}$  and $\varepsilon_{T,total}/[\kappa(\Delta_{T}/H)^{2}] \sim Re^{0.49}$, and the scaling exponents  are in excellent agreement with the global exact relation.
Compared with moderate-Prandtl number convection in the same cell, in the current low-Prandlt case the scaling exponent of $\varepsilon_{T,bulk} \sim Re^{0.57}$  is significantly larger; while the scaling exponent of $\varepsilon_{T,BL} \sim Re^{0.46}$  is slightly smaller.
\end{enumerate}

\section*{\label{SectionSM}supplementary material}

See the supplementary material for the video of instantaneous temperature and flow fields in both low- and moderate-Prandlt number turbulent thermal convection.

\begin{acknowledgments}
This work was supported by the National Natural Science Foundation of China (NSFC) through Grant Nos. 11902268 and 11772259, the Fundamental Research Funds for the Central Universities of China (Nos. G2019KY05101 and 3102019PJ002), and the 111 project of China (No. B17037).
The simulations were carried out at LvLiang Cloud Computing Center of China, and the calculations were performed on TianHe-2.
\end{acknowledgments}

\appendix*

\section{\label{appendix}Simulation settings for moderate-Prandtl number convection}

We simulated turbulent thermal convection at a moderate Prandtl number (i.e., $Pr = 5.3$ and $Ra = 10^{7}$) to compare with low-Prandtl number convection.
The mesh size was chosen as $257^{2}$, which resulted in $\Delta_{g}/\eta \approx 0.18$, $\Delta_{g}/\eta_{B} \approx 0.41$, and $\Delta_{t}/\tau_{\eta} \approx 0.00034$ (see Sec. \ref{subsectionSettings} for the definition of $\Delta_{g}$, $\eta$, $\eta_{B}$, $\Delta_{t}$ and $\tau_{\eta}$).
A total run-time of 1000 free-fall time units were adopted to obtain statistically convergent results.

\nocite{*}
\bibliography{myBib}% Produces the bibliography via BibTeX.

\begin{thebibliography}{52}%
\makeatletter
\providecommand \@ifxundefined [1]{%
 \@ifx{#1\undefined}
}%
\providecommand \@ifnum [1]{%
 \ifnum #1\expandafter \@firstoftwo
 \else \expandafter \@secondoftwo
 \fi
}%
\providecommand \@ifx [1]{%
 \ifx #1\expandafter \@firstoftwo
 \else \expandafter \@secondoftwo
 \fi
}%
\providecommand \natexlab [1]{#1}%
\providecommand \enquote  [1]{``#1''}%
\providecommand \bibnamefont  [1]{#1}%
\providecommand \bibfnamefont [1]{#1}%
\providecommand \citenamefont [1]{#1}%
\providecommand \href@noop [0]{\@secondoftwo}%
\providecommand \href [0]{\begingroup \@sanitize@url \@href}%
\providecommand \@href[1]{\@@startlink{#1}\@@href}%
\providecommand \@@href[1]{\endgroup#1\@@endlink}%
\providecommand \@sanitize@url [0]{\catcode `\\12\catcode `\$12\catcode
  `\&12\catcode `\#12\catcode `\^12\catcode `\_12\catcode `\%12\relax}%
\providecommand \@@startlink[1]{}%
\providecommand \@@endlink[0]{}%
\providecommand \url  [0]{\begingroup\@sanitize@url \@url }%
\providecommand \@url [1]{\endgroup\@href {#1}{\urlprefix }}%
\providecommand \urlprefix  [0]{URL }%
\providecommand \Eprint [0]{\href }%
\providecommand \doibase [0]{http://dx.doi.org/}%
\providecommand \selectlanguage [0]{\@gobble}%
\providecommand \bibinfo  [0]{\@secondoftwo}%
\providecommand \bibfield  [0]{\@secondoftwo}%
\providecommand \translation [1]{[#1]}%
\providecommand \BibitemOpen [0]{}%
\providecommand \bibitemStop [0]{}%
\providecommand \bibitemNoStop [0]{.\EOS\space}%
\providecommand \EOS [0]{\spacefactor3000\relax}%
\providecommand \BibitemShut  [1]{\csname bibitem#1\endcsname}%
\let\auto@bib@innerbib\@empty
%</preamble>
\bibitem [{\citenamefont {Hanasoge}, \citenamefont {Gizon},\ and\ \citenamefont
  {Sreenivasan}(2016)}]{hanasoge2016seismic}%
  \BibitemOpen
  \bibfield  {author} {\bibinfo {author} {\bibfnamefont {S.}~\bibnamefont
  {Hanasoge}}, \bibinfo {author} {\bibfnamefont {L.}~\bibnamefont {Gizon}}, \
  and\ \bibinfo {author} {\bibfnamefont {K.~R.}\ \bibnamefont {Sreenivasan}},\
  }\bibfield  {title} {\enquote {\bibinfo {title} {Seismic sounding of
  convection in the {S}un},}\ }\href {\doibase
  10.1146/annurev-fluid-122414-034534} {\bibfield  {journal} {\bibinfo
  {journal} {Annual Review of Fluid Mechanics}\ }\textbf {\bibinfo {volume}
  {48}},\ \bibinfo {pages} {191--217} (\bibinfo {year} {2016})}\BibitemShut
  {NoStop}%
\bibitem [{\citenamefont {King}\ and\ \citenamefont
  {Aurnou}(2013)}]{king2013turbulent}%
  \BibitemOpen
  \bibfield  {author} {\bibinfo {author} {\bibfnamefont {E.~M.}\ \bibnamefont
  {King}}\ and\ \bibinfo {author} {\bibfnamefont {J.~M.}\ \bibnamefont
  {Aurnou}},\ }\bibfield  {title} {\enquote {\bibinfo {title} {Turbulent
  convection in liquid metal with and without rotation},}\ }\href {\doibase
  10.1073/pnas.1217553110} {\bibfield  {journal} {\bibinfo  {journal}
  {Proceedings of the National Academy of Sciences}\ }\textbf {\bibinfo
  {volume} {110}},\ \bibinfo {pages} {6688--6693} (\bibinfo {year}
  {2013})}\BibitemShut {NoStop}%
\bibitem [{\citenamefont {Gr{\"o}tzbach}(2013)}]{grotzbach2013challenges}%
  \BibitemOpen
  \bibfield  {author} {\bibinfo {author} {\bibfnamefont {G.}~\bibnamefont
  {Gr{\"o}tzbach}},\ }\bibfield  {title} {\enquote {\bibinfo {title}
  {Challenges in low-{P}randtl number heat transfer simulation and
  modelling},}\ }\href {\doibase 10.1016/j.nucengdes.2012.09.039} {\bibfield
  {journal} {\bibinfo  {journal} {Nuclear Engineering and Design}\ }\textbf
  {\bibinfo {volume} {264}},\ \bibinfo {pages} {41--55} (\bibinfo {year}
  {2013})}\BibitemShut {NoStop}%
\bibitem [{\citenamefont {Kelley}\ and\ \citenamefont
  {Sadoway}(2014)}]{kelley2014mixing}%
  \BibitemOpen
  \bibfield  {author} {\bibinfo {author} {\bibfnamefont {D.~H.}\ \bibnamefont
  {Kelley}}\ and\ \bibinfo {author} {\bibfnamefont {D.~R.}\ \bibnamefont
  {Sadoway}},\ }\bibfield  {title} {\enquote {\bibinfo {title} {Mixing in a
  liquid metal electrode},}\ }\href {\doibase 10.1063/1.4875815} {\bibfield
  {journal} {\bibinfo  {journal} {Physics of Fluids}\ }\textbf {\bibinfo
  {volume} {26}},\ \bibinfo {pages} {057102} (\bibinfo {year}
  {2014})}\BibitemShut {NoStop}%
\bibitem [{\citenamefont {Ahlers}, \citenamefont {Grossmann},\ and\
  \citenamefont {Lohse}(2009)}]{ahlers2009heat}%
  \BibitemOpen
  \bibfield  {author} {\bibinfo {author} {\bibfnamefont {G.}~\bibnamefont
  {Ahlers}}, \bibinfo {author} {\bibfnamefont {S.}~\bibnamefont {Grossmann}}, \
  and\ \bibinfo {author} {\bibfnamefont {D.}~\bibnamefont {Lohse}},\ }\bibfield
   {title} {\enquote {\bibinfo {title} {Heat transfer and large scale dynamics
  in turbulent {R}ayleigh-{B}{\'e}nard convection},}\ }\href {\doibase
  10.1103/RevModPhys.81.503} {\bibfield  {journal} {\bibinfo  {journal}
  {Reviews of Modern Physics}\ }\textbf {\bibinfo {volume} {81}},\ \bibinfo
  {pages} {503} (\bibinfo {year} {2009})}\BibitemShut {NoStop}%
\bibitem [{\citenamefont {Lohse}\ and\ \citenamefont
  {Xia}(2010)}]{lohse2010small}%
  \BibitemOpen
  \bibfield  {author} {\bibinfo {author} {\bibfnamefont {D.}~\bibnamefont
  {Lohse}}\ and\ \bibinfo {author} {\bibfnamefont {K.-Q.}\ \bibnamefont
  {Xia}},\ }\bibfield  {title} {\enquote {\bibinfo {title} {Small-scale
  properties of turbulent {R}ayleigh-{B}{\'e}nard convection},}\ }\href@noop {}
  {\bibfield  {journal} {\bibinfo  {journal} {Annual Review of Fluid
  Mechanics}\ }\textbf {\bibinfo {volume} {42}} (\bibinfo {year}
  {2010})}\BibitemShut {NoStop}%
\bibitem [{\citenamefont {Chill{\`a}}\ and\ \citenamefont
  {Schumacher}(2012)}]{chilla2012new}%
  \BibitemOpen
  \bibfield  {author} {\bibinfo {author} {\bibfnamefont {F.}~\bibnamefont
  {Chill{\`a}}}\ and\ \bibinfo {author} {\bibfnamefont {J.}~\bibnamefont
  {Schumacher}},\ }\bibfield  {title} {\enquote {\bibinfo {title} {New
  perspectives in turbulent {R}ayleigh-{B}{\'e}nard convection},}\ }\href
  {\doibase 10.1140/epje/i2012-12058-1} {\bibfield  {journal} {\bibinfo
  {journal} {The European Physical Journal E}\ }\textbf {\bibinfo {volume}
  {35}},\ \bibinfo {pages} {58} (\bibinfo {year} {2012})}\BibitemShut {NoStop}%
\bibitem [{\citenamefont {Xia}(2013)}]{xia2013current}%
  \BibitemOpen
  \bibfield  {author} {\bibinfo {author} {\bibfnamefont {K.-Q.}\ \bibnamefont
  {Xia}},\ }\bibfield  {title} {\enquote {\bibinfo {title} {Current trends and
  future directions in turbulent thermal convection},}\ }\href {\doibase
  10.1063/2.1305201} {\bibfield  {journal} {\bibinfo  {journal} {Theoretical
  and Applied Mechanics Letters}\ }\textbf {\bibinfo {volume} {3}},\ \bibinfo
  {pages} {052001} (\bibinfo {year} {2013})}\BibitemShut {NoStop}%
\bibitem [{\citenamefont {Mazzino}(2017)}]{mazzino2017two}%
  \BibitemOpen
  \bibfield  {author} {\bibinfo {author} {\bibfnamefont {A.}~\bibnamefont
  {Mazzino}},\ }\bibfield  {title} {\enquote {\bibinfo {title} {Two-dimensional
  turbulent convection},}\ }\href {\doibase 10.1063/1.4990083} {\bibfield
  {journal} {\bibinfo  {journal} {Physics of Fluids}\ }\textbf {\bibinfo
  {volume} {29}},\ \bibinfo {pages} {111102} (\bibinfo {year}
  {2017})}\BibitemShut {NoStop}%
\bibitem [{\citenamefont {Vogt}\ \emph {et~al.}(2018)\citenamefont {Vogt},
  \citenamefont {Horn}, \citenamefont {Grannan},\ and\ \citenamefont
  {Aurnou}}]{vogt2018jump}%
  \BibitemOpen
  \bibfield  {author} {\bibinfo {author} {\bibfnamefont {T.}~\bibnamefont
  {Vogt}}, \bibinfo {author} {\bibfnamefont {S.}~\bibnamefont {Horn}}, \bibinfo
  {author} {\bibfnamefont {A.~M.}\ \bibnamefont {Grannan}}, \ and\ \bibinfo
  {author} {\bibfnamefont {J.~M.}\ \bibnamefont {Aurnou}},\ }\bibfield  {title}
  {\enquote {\bibinfo {title} {Jump rope vortex in liquid metal convection},}\
  }\href {\doibase 10.1073/pnas.1812260115} {\bibfield  {journal} {\bibinfo
  {journal} {Proceedings of the National Academy of Sciences}\ }\textbf
  {\bibinfo {volume} {115}},\ \bibinfo {pages} {12674--12679} (\bibinfo {year}
  {2018})}\BibitemShut {NoStop}%
\bibitem [{\citenamefont {Schumacher}, \citenamefont {G{\"o}tzfried},\ and\
  \citenamefont {Scheel}(2015)}]{schumacher2015enhanced}%
  \BibitemOpen
  \bibfield  {author} {\bibinfo {author} {\bibfnamefont {J.}~\bibnamefont
  {Schumacher}}, \bibinfo {author} {\bibfnamefont {P.}~\bibnamefont
  {G{\"o}tzfried}}, \ and\ \bibinfo {author} {\bibfnamefont {J.~D.}\
  \bibnamefont {Scheel}},\ }\bibfield  {title} {\enquote {\bibinfo {title}
  {Enhanced enstrophy generation for turbulent convection in
  low-{P}randtl-number fluids},}\ }\href {\doibase 10.1073/pnas.1505111112}
  {\bibfield  {journal} {\bibinfo  {journal} {Proceedings of the National
  Academy of Sciences}\ }\textbf {\bibinfo {volume} {112}},\ \bibinfo {pages}
  {9530--9535} (\bibinfo {year} {2015})}\BibitemShut {NoStop}%
\bibitem [{\citenamefont {Scheel}\ and\ \citenamefont
  {Schumacher}(2016)}]{scheel2016global}%
  \BibitemOpen
  \bibfield  {author} {\bibinfo {author} {\bibfnamefont {J.~D.}\ \bibnamefont
  {Scheel}}\ and\ \bibinfo {author} {\bibfnamefont {J.}~\bibnamefont
  {Schumacher}},\ }\bibfield  {title} {\enquote {\bibinfo {title} {Global and
  local statistics in turbulent convection at low {P}randtl numbers},}\ }\href
  {\doibase 10.1017/jfm.2016.457} {\bibfield  {journal} {\bibinfo  {journal}
  {Journal of Fluid Mechanics}\ }\textbf {\bibinfo {volume} {802}},\ \bibinfo
  {pages} {147--173} (\bibinfo {year} {2016})}\BibitemShut {NoStop}%
\bibitem [{\citenamefont {Zwirner}\ and\ \citenamefont
  {Shishkina}(2018)}]{zwirner2018confined}%
  \BibitemOpen
  \bibfield  {author} {\bibinfo {author} {\bibfnamefont {L.}~\bibnamefont
  {Zwirner}}\ and\ \bibinfo {author} {\bibfnamefont {O.}~\bibnamefont
  {Shishkina}},\ }\bibfield  {title} {\enquote {\bibinfo {title} {Confined
  inclined thermal convection in low-{P}randtl-number fluids},}\ }\href
  {\doibase 10.1017/jfm.2018.477} {\bibfield  {journal} {\bibinfo  {journal}
  {Journal of Fluid Mechanics}\ }\textbf {\bibinfo {volume} {850}},\ \bibinfo
  {pages} {984--1008} (\bibinfo {year} {2018})}\BibitemShut {NoStop}%
\bibitem [{\citenamefont {Shraiman}\ and\ \citenamefont
  {Siggia}(1990)}]{shraiman1990heat}%
  \BibitemOpen
  \bibfield  {author} {\bibinfo {author} {\bibfnamefont {B.~I.}\ \bibnamefont
  {Shraiman}}\ and\ \bibinfo {author} {\bibfnamefont {E.~D.}\ \bibnamefont
  {Siggia}},\ }\bibfield  {title} {\enquote {\bibinfo {title} {Heat transport
  in high-{R}ayleigh-number convection},}\ }\href {\doibase
  10.1103/PhysRevA.42.3650} {\bibfield  {journal} {\bibinfo  {journal}
  {Physical Review A}\ }\textbf {\bibinfo {volume} {42}},\ \bibinfo {pages}
  {3650} (\bibinfo {year} {1990})}\BibitemShut {NoStop}%
\bibitem [{\citenamefont {Grossmann}\ and\ \citenamefont
  {Lohse}(2000)}]{grossmann2000scaling}%
  \BibitemOpen
  \bibfield  {author} {\bibinfo {author} {\bibfnamefont {S.}~\bibnamefont
  {Grossmann}}\ and\ \bibinfo {author} {\bibfnamefont {D.}~\bibnamefont
  {Lohse}},\ }\bibfield  {title} {\enquote {\bibinfo {title} {Scaling in
  thermal convection: a unifying theory},}\ }\href {\doibase
  10.1017/S0022112099007545} {\bibfield  {journal} {\bibinfo  {journal}
  {Journal of Fluid Mechanics}\ }\textbf {\bibinfo {volume} {407}},\ \bibinfo
  {pages} {27--56} (\bibinfo {year} {2000})}\BibitemShut {NoStop}%
\bibitem [{\citenamefont {Grossmann}\ and\ \citenamefont
  {Lohse}(2002)}]{grossmann2002prandtl}%
  \BibitemOpen
  \bibfield  {author} {\bibinfo {author} {\bibfnamefont {S.}~\bibnamefont
  {Grossmann}}\ and\ \bibinfo {author} {\bibfnamefont {D.}~\bibnamefont
  {Lohse}},\ }\bibfield  {title} {\enquote {\bibinfo {title} {Prandtl and
  {R}ayleigh number dependence of the {R}eynolds number in turbulent thermal
  convection},}\ }\href {\doibase 10.1103/PhysRevE.66.016305} {\bibfield
  {journal} {\bibinfo  {journal} {Physical Review E}\ }\textbf {\bibinfo
  {volume} {66}},\ \bibinfo {pages} {016305} (\bibinfo {year}
  {2002})}\BibitemShut {NoStop}%
\bibitem [{\citenamefont {Grossmann}\ and\ \citenamefont
  {Lohse}(2004)}]{grossmann2004fluctuations}%
  \BibitemOpen
  \bibfield  {author} {\bibinfo {author} {\bibfnamefont {S.}~\bibnamefont
  {Grossmann}}\ and\ \bibinfo {author} {\bibfnamefont {D.}~\bibnamefont
  {Lohse}},\ }\bibfield  {title} {\enquote {\bibinfo {title} {Fluctuations in
  turbulent {R}ayleigh--{B}{\'e}nard convection: the role of plumes},}\ }\href
  {\doibase 10.1063/1.1807751} {\bibfield  {journal} {\bibinfo  {journal}
  {Physics of Fluids}\ }\textbf {\bibinfo {volume} {16}},\ \bibinfo {pages}
  {4462--4472} (\bibinfo {year} {2004})}\BibitemShut {NoStop}%
\bibitem [{\citenamefont {Emran}\ and\ \citenamefont
  {Schumacher}(2008)}]{emran2008fine}%
  \BibitemOpen
  \bibfield  {author} {\bibinfo {author} {\bibfnamefont {M.}~\bibnamefont
  {Emran}}\ and\ \bibinfo {author} {\bibfnamefont {J.}~\bibnamefont
  {Schumacher}},\ }\bibfield  {title} {\enquote {\bibinfo {title} {Fine-scale
  statistics of temperature and its derivatives in convective turbulence},}\
  }\href {\doibase 10.1017/S0022112008002954} {\bibfield  {journal} {\bibinfo
  {journal} {Journal of Fluid Mechanics}\ }\textbf {\bibinfo {volume} {611}},\
  \bibinfo {pages} {13--34} (\bibinfo {year} {2008})}\BibitemShut {NoStop}%
\bibitem [{\citenamefont {Overholt}\ and\ \citenamefont
  {Pope}(1996)}]{overholt1996direct}%
  \BibitemOpen
  \bibfield  {author} {\bibinfo {author} {\bibfnamefont {M.}~\bibnamefont
  {Overholt}}\ and\ \bibinfo {author} {\bibfnamefont {S.}~\bibnamefont
  {Pope}},\ }\bibfield  {title} {\enquote {\bibinfo {title} {Direct numerical
  simulation of a passive scalar with imposed mean gradient in isotropic
  turbulence},}\ }\href {\doibase 10.1063/1.869099} {\bibfield  {journal}
  {\bibinfo  {journal} {Physics of Fluids}\ }\textbf {\bibinfo {volume} {8}},\
  \bibinfo {pages} {3128--3148} (\bibinfo {year} {1996})}\BibitemShut {NoStop}%
\bibitem [{\citenamefont {Kaczorowski}\ and\ \citenamefont
  {Wagner}(2009)}]{kaczorowski2009analysis}%
  \BibitemOpen
  \bibfield  {author} {\bibinfo {author} {\bibfnamefont {M.}~\bibnamefont
  {Kaczorowski}}\ and\ \bibinfo {author} {\bibfnamefont {C.}~\bibnamefont
  {Wagner}},\ }\bibfield  {title} {\enquote {\bibinfo {title} {Analysis of the
  thermal plumes in turbulent {R}ayleigh--{B}{\'e}nard convection based on
  well-resolved numerical simulations},}\ }\href {\doibase
  10.1017/S0022112008003947} {\bibfield  {journal} {\bibinfo  {journal}
  {Journal of Fluid Mechanics}\ }\textbf {\bibinfo {volume} {618}},\ \bibinfo
  {pages} {89--112} (\bibinfo {year} {2009})}\BibitemShut {NoStop}%
\bibitem [{\citenamefont {Zhang}, \citenamefont {Zhou},\ and\ \citenamefont
  {Sun}(2017)}]{zhang2017statisticsJFM}%
  \BibitemOpen
  \bibfield  {author} {\bibinfo {author} {\bibfnamefont {Y.}~\bibnamefont
  {Zhang}}, \bibinfo {author} {\bibfnamefont {Q.}~\bibnamefont {Zhou}}, \ and\
  \bibinfo {author} {\bibfnamefont {C.}~\bibnamefont {Sun}},\ }\bibfield
  {title} {\enquote {\bibinfo {title} {Statistics of kinetic and thermal energy
  dissipation rates in two-dimensional turbulent {R}ayleigh--{B}{\'e}nard
  convection},}\ }\href {\doibase 10.1017/jfm.2017.19} {\bibfield  {journal}
  {\bibinfo  {journal} {Journal of Fluid Mechanics}\ }\textbf {\bibinfo
  {volume} {814}},\ \bibinfo {pages} {165--184} (\bibinfo {year}
  {2017})}\BibitemShut {NoStop}%
\bibitem [{\citenamefont {Zhang}\ \emph {et~al.}(2017)\citenamefont {Zhang},
  \citenamefont {Huang}, \citenamefont {Jiang}, \citenamefont {Liu},
  \citenamefont {Lu}, \citenamefont {Qiu},\ and\ \citenamefont
  {Zhou}}]{zhang2017statisticsPRE}%
  \BibitemOpen
  \bibfield  {author} {\bibinfo {author} {\bibfnamefont {Y.}~\bibnamefont
  {Zhang}}, \bibinfo {author} {\bibfnamefont {Y.-X.}\ \bibnamefont {Huang}},
  \bibinfo {author} {\bibfnamefont {N.}~\bibnamefont {Jiang}}, \bibinfo
  {author} {\bibfnamefont {Y.-L.}\ \bibnamefont {Liu}}, \bibinfo {author}
  {\bibfnamefont {Z.-M.}\ \bibnamefont {Lu}}, \bibinfo {author} {\bibfnamefont
  {X.}~\bibnamefont {Qiu}}, \ and\ \bibinfo {author} {\bibfnamefont
  {Q.}~\bibnamefont {Zhou}},\ }\bibfield  {title} {\enquote {\bibinfo {title}
  {Statistics of velocity and temperature fluctuations in two-dimensional
  {R}ayleigh-{B}{\'e}nard convection},}\ }\href {\doibase
  10.1103/PhysRevE.96.023105} {\bibfield  {journal} {\bibinfo  {journal}
  {Physical Review E}\ }\textbf {\bibinfo {volume} {96}},\ \bibinfo {pages}
  {023105} (\bibinfo {year} {2017})}\BibitemShut {NoStop}%
\bibitem [{\citenamefont {Bhattacharya}\ \emph {et~al.}(2018)\citenamefont
  {Bhattacharya}, \citenamefont {Pandey}, \citenamefont {Kumar},\ and\
  \citenamefont {Verma}}]{bhattacharya2018complexity}%
  \BibitemOpen
  \bibfield  {author} {\bibinfo {author} {\bibfnamefont {S.}~\bibnamefont
  {Bhattacharya}}, \bibinfo {author} {\bibfnamefont {A.}~\bibnamefont
  {Pandey}}, \bibinfo {author} {\bibfnamefont {A.}~\bibnamefont {Kumar}}, \
  and\ \bibinfo {author} {\bibfnamefont {M.~K.}\ \bibnamefont {Verma}},\
  }\bibfield  {title} {\enquote {\bibinfo {title} {Complexity of viscous
  dissipation in turbulent thermal convection},}\ }\href {\doibase
  10.1063/1.5022316} {\bibfield  {journal} {\bibinfo  {journal} {Physics of
  Fluids}\ }\textbf {\bibinfo {volume} {30}},\ \bibinfo {pages} {031702}
  (\bibinfo {year} {2018})}\BibitemShut {NoStop}%
\bibitem [{\citenamefont {Chen}\ and\ \citenamefont
  {Doolen}(1998)}]{chen1998lattice}%
  \BibitemOpen
  \bibfield  {author} {\bibinfo {author} {\bibfnamefont {S.}~\bibnamefont
  {Chen}}\ and\ \bibinfo {author} {\bibfnamefont {G.~D.}\ \bibnamefont
  {Doolen}},\ }\bibfield  {title} {\enquote {\bibinfo {title} {Lattice
  {B}oltzmann method for fluid flows},}\ }\href {\doibase
  10.1146/annurev.fluid.30.1.329} {\bibfield  {journal} {\bibinfo  {journal}
  {Annual Review of Fluid Mechanics}\ }\textbf {\bibinfo {volume} {30}},\
  \bibinfo {pages} {329--364} (\bibinfo {year} {1998})}\BibitemShut {NoStop}%
\bibitem [{\citenamefont {Aidun}\ and\ \citenamefont
  {Clausen}(2010)}]{aidun2010lattice}%
  \BibitemOpen
  \bibfield  {author} {\bibinfo {author} {\bibfnamefont {C.~K.}\ \bibnamefont
  {Aidun}}\ and\ \bibinfo {author} {\bibfnamefont {J.~R.}\ \bibnamefont
  {Clausen}},\ }\bibfield  {title} {\enquote {\bibinfo {title}
  {Lattice-{B}oltzmann method for complex flows},}\ }\href {\doibase
  10.1146/annurev-fluid-121108-145519} {\bibfield  {journal} {\bibinfo
  {journal} {Annual Review of Fluid Mechanics}\ }\textbf {\bibinfo {volume}
  {42}},\ \bibinfo {pages} {439--472} (\bibinfo {year} {2010})}\BibitemShut
  {NoStop}%
\bibitem [{\citenamefont {Xu}, \citenamefont {Shyy},\ and\ \citenamefont
  {Zhao}(2017)}]{xu2017lattice}%
  \BibitemOpen
  \bibfield  {author} {\bibinfo {author} {\bibfnamefont {A.}~\bibnamefont
  {Xu}}, \bibinfo {author} {\bibfnamefont {W.}~\bibnamefont {Shyy}}, \ and\
  \bibinfo {author} {\bibfnamefont {T.}~\bibnamefont {Zhao}},\ }\bibfield
  {title} {\enquote {\bibinfo {title} {Lattice {B}oltzmann modeling of
  transport phenomena in fuel cells and flow batteries},}\ }\href {\doibase
  10.1007/s10409-017-0667-6} {\bibfield  {journal} {\bibinfo  {journal} {Acta
  Mechanica Sinica}\ }\textbf {\bibinfo {volume} {33}},\ \bibinfo {pages}
  {555--574} (\bibinfo {year} {2017})}\BibitemShut {NoStop}%
\bibitem [{\citenamefont {Yu}, \citenamefont {Girimaji},\ and\ \citenamefont
  {Luo}(2005)}]{yu2005dns}%
  \BibitemOpen
  \bibfield  {author} {\bibinfo {author} {\bibfnamefont {H.}~\bibnamefont
  {Yu}}, \bibinfo {author} {\bibfnamefont {S.~S.}\ \bibnamefont {Girimaji}}, \
  and\ \bibinfo {author} {\bibfnamefont {L.-S.}\ \bibnamefont {Luo}},\
  }\bibfield  {title} {\enquote {\bibinfo {title} {{DNS} and {LES} of decaying
  isotropic turbulence with and without frame rotation using lattice
  {B}oltzmann method},}\ }\href {\doibase 10.1016/j.jcp.2005.03.022} {\bibfield
   {journal} {\bibinfo  {journal} {Journal of Computational Physics}\ }\textbf
  {\bibinfo {volume} {209}},\ \bibinfo {pages} {599--616} (\bibinfo {year}
  {2005})}\BibitemShut {NoStop}%
\bibitem [{\citenamefont {Wang}, \citenamefont {Wang},\ and\ \citenamefont
  {Guo}(2016)}]{wang2016comparison}%
  \BibitemOpen
  \bibfield  {author} {\bibinfo {author} {\bibfnamefont {P.}~\bibnamefont
  {Wang}}, \bibinfo {author} {\bibfnamefont {L.-P.}\ \bibnamefont {Wang}}, \
  and\ \bibinfo {author} {\bibfnamefont {Z.}~\bibnamefont {Guo}},\ }\bibfield
  {title} {\enquote {\bibinfo {title} {Comparison of the lattice {B}oltzmann
  equation and discrete unified gas-kinetic scheme methods for direct numerical
  simulation of decaying turbulent flows},}\ }\href {\doibase
  10.1103/PhysRevE.94.043304} {\bibfield  {journal} {\bibinfo  {journal}
  {Physical Review E}\ }\textbf {\bibinfo {volume} {94}},\ \bibinfo {pages}
  {043304} (\bibinfo {year} {2016})}\BibitemShut {NoStop}%
\bibitem [{\citenamefont {Peng}\ \emph {et~al.}(2018)\citenamefont {Peng},
  \citenamefont {Geneva}, \citenamefont {Guo},\ and\ \citenamefont
  {Wang}}]{peng2018direct}%
  \BibitemOpen
  \bibfield  {author} {\bibinfo {author} {\bibfnamefont {C.}~\bibnamefont
  {Peng}}, \bibinfo {author} {\bibfnamefont {N.}~\bibnamefont {Geneva}},
  \bibinfo {author} {\bibfnamefont {Z.}~\bibnamefont {Guo}}, \ and\ \bibinfo
  {author} {\bibfnamefont {L.-P.}\ \bibnamefont {Wang}},\ }\bibfield  {title}
  {\enquote {\bibinfo {title} {Direct numerical simulation of turbulent pipe
  flow using the lattice {B}oltzmann method},}\ }\href {\doibase
  10.1016/j.jcp.2017.11.040} {\bibfield  {journal} {\bibinfo  {journal}
  {Journal of Computational Physics}\ }\textbf {\bibinfo {volume} {357}},\
  \bibinfo {pages} {16--42} (\bibinfo {year} {2018})}\BibitemShut {NoStop}%
\bibitem [{\citenamefont {Peng}, \citenamefont {Ayala},\ and\ \citenamefont
  {Wang}(2019)}]{peng2019directJFM}%
  \BibitemOpen
  \bibfield  {author} {\bibinfo {author} {\bibfnamefont {C.}~\bibnamefont
  {Peng}}, \bibinfo {author} {\bibfnamefont {O.~M.}\ \bibnamefont {Ayala}}, \
  and\ \bibinfo {author} {\bibfnamefont {L.-P.}\ \bibnamefont {Wang}},\
  }\bibfield  {title} {\enquote {\bibinfo {title} {A direct numerical
  investigation of two-way interactions in a particle-laden turbulent channel
  flow},}\ }\href {\doibase 10.1017/jfm.2019.509} {\bibfield  {journal}
  {\bibinfo  {journal} {Journal of Fluid Mechanics}\ }\textbf {\bibinfo
  {volume} {875}},\ \bibinfo {pages} {1096--1144} (\bibinfo {year}
  {2019})}\BibitemShut {NoStop}%
\bibitem [{\citenamefont {Xu}, \citenamefont {Shi},\ and\ \citenamefont
  {Xi}(2019)}]{xu2019lattice}%
  \BibitemOpen
  \bibfield  {author} {\bibinfo {author} {\bibfnamefont {A.}~\bibnamefont
  {Xu}}, \bibinfo {author} {\bibfnamefont {L.}~\bibnamefont {Shi}}, \ and\
  \bibinfo {author} {\bibfnamefont {H.-D.}\ \bibnamefont {Xi}},\ }\bibfield
  {title} {\enquote {\bibinfo {title} {Lattice {B}oltzmann simulations of
  three-dimensional thermal convective flows at high {R}ayleigh number},}\
  }\href {\doibase 10.1016/j.ijheatmasstransfer.2019.06.002} {\bibfield
  {journal} {\bibinfo  {journal} {International Journal of Heat and Mass
  Transfer}\ }\textbf {\bibinfo {volume} {140}},\ \bibinfo {pages} {359--370}
  (\bibinfo {year} {2019})}\BibitemShut {NoStop}%
\bibitem [{\citenamefont {Guo}, \citenamefont {Zheng},\ and\ \citenamefont
  {Shi}(2002)}]{guo2002discrete}%
  \BibitemOpen
  \bibfield  {author} {\bibinfo {author} {\bibfnamefont {Z.}~\bibnamefont
  {Guo}}, \bibinfo {author} {\bibfnamefont {C.}~\bibnamefont {Zheng}}, \ and\
  \bibinfo {author} {\bibfnamefont {B.}~\bibnamefont {Shi}},\ }\bibfield
  {title} {\enquote {\bibinfo {title} {Discrete lattice effects on the forcing
  term in the lattice {B}oltzmann method},}\ }\href {\doibase
  10.1103/PhysRevE.65.046308} {\bibfield  {journal} {\bibinfo  {journal}
  {Physical Review E}\ }\textbf {\bibinfo {volume} {65}},\ \bibinfo {pages}
  {046308} (\bibinfo {year} {2002})}\BibitemShut {NoStop}%
\bibitem [{\citenamefont {Dubois}\ and\ \citenamefont
  {Lallemand}(2009)}]{dubois2009towards}%
  \BibitemOpen
  \bibfield  {author} {\bibinfo {author} {\bibfnamefont {F.}~\bibnamefont
  {Dubois}}\ and\ \bibinfo {author} {\bibfnamefont {P.}~\bibnamefont
  {Lallemand}},\ }\bibfield  {title} {\enquote {\bibinfo {title} {Towards
  higher order lattice {B}oltzmann schemes},}\ }\href {\doibase
  10.1088/1742-5468/2009/06/P06006} {\bibfield  {journal} {\bibinfo  {journal}
  {Journal of Statistical Mechanics: Theory and Experiment}\ }\textbf {\bibinfo
  {volume} {2009}},\ \bibinfo {pages} {P06006} (\bibinfo {year}
  {2009})}\BibitemShut {NoStop}%
\bibitem [{\citenamefont {Wang}\ \emph {et~al.}(2013)\citenamefont {Wang},
  \citenamefont {Wang}, \citenamefont {Lallemand},\ and\ \citenamefont
  {Luo}}]{wang2013lattice}%
  \BibitemOpen
  \bibfield  {author} {\bibinfo {author} {\bibfnamefont {J.}~\bibnamefont
  {Wang}}, \bibinfo {author} {\bibfnamefont {D.}~\bibnamefont {Wang}}, \bibinfo
  {author} {\bibfnamefont {P.}~\bibnamefont {Lallemand}}, \ and\ \bibinfo
  {author} {\bibfnamefont {L.-S.}\ \bibnamefont {Luo}},\ }\bibfield  {title}
  {\enquote {\bibinfo {title} {Lattice {B}oltzmann simulations of thermal
  convective flows in two dimensions},}\ }\href {\doibase
  10.1016/j.camwa.2012.07.001} {\bibfield  {journal} {\bibinfo  {journal}
  {Computers \& Mathematics with Applications}\ }\textbf {\bibinfo {volume}
  {65}},\ \bibinfo {pages} {262--286} (\bibinfo {year} {2013})}\BibitemShut
  {NoStop}%
\bibitem [{\citenamefont {Contrino}\ \emph {et~al.}(2014)\citenamefont
  {Contrino}, \citenamefont {Lallemand}, \citenamefont {Asinari},\ and\
  \citenamefont {Luo}}]{contrino2014lattice}%
  \BibitemOpen
  \bibfield  {author} {\bibinfo {author} {\bibfnamefont {D.}~\bibnamefont
  {Contrino}}, \bibinfo {author} {\bibfnamefont {P.}~\bibnamefont {Lallemand}},
  \bibinfo {author} {\bibfnamefont {P.}~\bibnamefont {Asinari}}, \ and\
  \bibinfo {author} {\bibfnamefont {L.-S.}\ \bibnamefont {Luo}},\ }\bibfield
  {title} {\enquote {\bibinfo {title} {Lattice-{B}oltzmann simulations of the
  thermally driven 2{D} square cavity at high {R}ayleigh numbers},}\ }\href
  {\doibase 10.1016/j.jcp.2014.06.047} {\bibfield  {journal} {\bibinfo
  {journal} {Journal of Computational Physics}\ }\textbf {\bibinfo {volume}
  {275}},\ \bibinfo {pages} {257--272} (\bibinfo {year} {2014})}\BibitemShut
  {NoStop}%
\bibitem [{\citenamefont {Xu}, \citenamefont {Shi},\ and\ \citenamefont
  {Zhao}(2017)}]{xu2017accelerated}%
  \BibitemOpen
  \bibfield  {author} {\bibinfo {author} {\bibfnamefont {A.}~\bibnamefont
  {Xu}}, \bibinfo {author} {\bibfnamefont {L.}~\bibnamefont {Shi}}, \ and\
  \bibinfo {author} {\bibfnamefont {T.}~\bibnamefont {Zhao}},\ }\bibfield
  {title} {\enquote {\bibinfo {title} {Accelerated lattice {B}oltzmann
  simulation using {GPU} and {O}pen{ACC} with data management},}\ }\href
  {\doibase 10.1016/j.ijheatmasstransfer.2017.02.032} {\bibfield  {journal}
  {\bibinfo  {journal} {International Journal of Heat and Mass Transfer}\
  }\textbf {\bibinfo {volume} {109}},\ \bibinfo {pages} {577--588} (\bibinfo
  {year} {2017})}\BibitemShut {NoStop}%
\bibitem [{\citenamefont {Kerr}(1996)}]{kerr1996rayleigh}%
  \BibitemOpen
  \bibfield  {author} {\bibinfo {author} {\bibfnamefont {R.~M.}\ \bibnamefont
  {Kerr}},\ }\bibfield  {title} {\enquote {\bibinfo {title} {Rayleigh number
  scaling in numerical convection},}\ }\href {\doibase
  10.1017/S0022112096001760} {\bibfield  {journal} {\bibinfo  {journal}
  {Journal of Fluid Mechanics}\ }\textbf {\bibinfo {volume} {310}},\ \bibinfo
  {pages} {139--179} (\bibinfo {year} {1996})}\BibitemShut {NoStop}%
\bibitem [{\citenamefont {Shishkina}\ and\ \citenamefont
  {Wagner}(2007)}]{shishkina2007local}%
  \BibitemOpen
  \bibfield  {author} {\bibinfo {author} {\bibfnamefont {O.}~\bibnamefont
  {Shishkina}}\ and\ \bibinfo {author} {\bibfnamefont {C.}~\bibnamefont
  {Wagner}},\ }\bibfield  {title} {\enquote {\bibinfo {title} {Local heat
  fluxes in turbulent {R}ayleigh-{B}{\'e}nard convection},}\ }\href {\doibase
  10.1063/1.2756583} {\bibfield  {journal} {\bibinfo  {journal} {Physics of
  Fluids}\ }\textbf {\bibinfo {volume} {19}},\ \bibinfo {pages} {085107}
  (\bibinfo {year} {2007})}\BibitemShut {NoStop}%
\bibitem [{\citenamefont {Zhou}\ and\ \citenamefont
  {Chen}(2018)}]{zhou2018similarity}%
  \BibitemOpen
  \bibfield  {author} {\bibinfo {author} {\bibfnamefont {W.-F.}\ \bibnamefont
  {Zhou}}\ and\ \bibinfo {author} {\bibfnamefont {J.}~\bibnamefont {Chen}},\
  }\bibfield  {title} {\enquote {\bibinfo {title} {Letter: {S}imilarity model
  for corner roll in turbulent {R}ayleigh-{B}{\'e}nard convection},}\ }\href
  {\doibase 10.1063/1.5054647} {\bibfield  {journal} {\bibinfo  {journal}
  {Physics of Fluids}\ }\textbf {\bibinfo {volume} {30}},\ \bibinfo {pages}
  {111705} (\bibinfo {year} {2018})}\BibitemShut {NoStop}%
\bibitem [{\citenamefont {Cioni}, \citenamefont {Ciliberto},\ and\
  \citenamefont {Sommeria}(1997)}]{cioni1997strongly}%
  \BibitemOpen
  \bibfield  {author} {\bibinfo {author} {\bibfnamefont {S.}~\bibnamefont
  {Cioni}}, \bibinfo {author} {\bibfnamefont {S.}~\bibnamefont {Ciliberto}}, \
  and\ \bibinfo {author} {\bibfnamefont {J.}~\bibnamefont {Sommeria}},\
  }\bibfield  {title} {\enquote {\bibinfo {title} {Strongly turbulent
  {R}ayleigh--{B}{\'e}nard convection in mercury: comparison with results at
  moderate {P}randtl number},}\ }\href {\doibase 10.1017/S0022112096004491}
  {\bibfield  {journal} {\bibinfo  {journal} {Journal of Fluid Mechanics}\
  }\textbf {\bibinfo {volume} {335}},\ \bibinfo {pages} {111--140} (\bibinfo
  {year} {1997})}\BibitemShut {NoStop}%
\bibitem [{\citenamefont {Scheel}\ and\ \citenamefont
  {Schumacher}(2017)}]{scheel2017predicting}%
  \BibitemOpen
  \bibfield  {author} {\bibinfo {author} {\bibfnamefont {J.~D.}\ \bibnamefont
  {Scheel}}\ and\ \bibinfo {author} {\bibfnamefont {J.}~\bibnamefont
  {Schumacher}},\ }\bibfield  {title} {\enquote {\bibinfo {title} {Predicting
  transition ranges to fully turbulent viscous boundary layers in low {P}randtl
  number convection flows},}\ }\href {\doibase 10.1103/PhysRevFluids.2.123501}
  {\bibfield  {journal} {\bibinfo  {journal} {Physical Review Fluids}\ }\textbf
  {\bibinfo {volume} {2}},\ \bibinfo {pages} {123501} (\bibinfo {year}
  {2017})}\BibitemShut {NoStop}%
\bibitem [{\citenamefont {van~der Poel}, \citenamefont {Stevens},\ and\
  \citenamefont {Lohse}(2013)}]{van2013comparison}%
  \BibitemOpen
  \bibfield  {author} {\bibinfo {author} {\bibfnamefont {E.~P.}\ \bibnamefont
  {van~der Poel}}, \bibinfo {author} {\bibfnamefont {R.~J.}\ \bibnamefont
  {Stevens}}, \ and\ \bibinfo {author} {\bibfnamefont {D.}~\bibnamefont
  {Lohse}},\ }\bibfield  {title} {\enquote {\bibinfo {title} {Comparison
  between two-and three-dimensional {R}ayleigh--{B}{\'e}nard convection},}\
  }\href {\doibase 10.1017/jfm.2013.488} {\bibfield  {journal} {\bibinfo
  {journal} {Journal of Fluid Mechanics}\ }\textbf {\bibinfo {volume} {736}},\
  \bibinfo {pages} {177--194} (\bibinfo {year} {2013})}\BibitemShut {NoStop}%
\bibitem [{\citenamefont {Zhang}\ \emph {et~al.}(2018)\citenamefont {Zhang},
  \citenamefont {Sun}, \citenamefont {Bao},\ and\ \citenamefont
  {Zhou}}]{zhang2018surface}%
  \BibitemOpen
  \bibfield  {author} {\bibinfo {author} {\bibfnamefont {Y.-Z.}\ \bibnamefont
  {Zhang}}, \bibinfo {author} {\bibfnamefont {C.}~\bibnamefont {Sun}}, \bibinfo
  {author} {\bibfnamefont {Y.}~\bibnamefont {Bao}}, \ and\ \bibinfo {author}
  {\bibfnamefont {Q.}~\bibnamefont {Zhou}},\ }\bibfield  {title} {\enquote
  {\bibinfo {title} {How surface roughness reduces heat transport for small
  roughness heights in turbulent {R}ayleigh--{B}{\'e}nard convection},}\ }\href
  {\doibase 10.1017/jfm.2017.786} {\bibfield  {journal} {\bibinfo  {journal}
  {Journal of Fluid Mechanics}\ }\textbf {\bibinfo {volume} {836}} (\bibinfo
  {year} {2018}),\ 10.1017/jfm.2017.786}\BibitemShut {NoStop}%
\bibitem [{\citenamefont {Johnston}\ and\ \citenamefont
  {Doering}(2009)}]{johnston2009comparison}%
  \BibitemOpen
  \bibfield  {author} {\bibinfo {author} {\bibfnamefont {H.}~\bibnamefont
  {Johnston}}\ and\ \bibinfo {author} {\bibfnamefont {C.~R.}\ \bibnamefont
  {Doering}},\ }\bibfield  {title} {\enquote {\bibinfo {title} {Comparison of
  turbulent thermal convection between conditions of constant temperature and
  constant flux},}\ }\href {\doibase 10.1103/PhysRevLett.102.064501} {\bibfield
   {journal} {\bibinfo  {journal} {Physical Review Letters}\ }\textbf {\bibinfo
  {volume} {102}},\ \bibinfo {pages} {064501} (\bibinfo {year}
  {2009})}\BibitemShut {NoStop}%
\bibitem [{\citenamefont {van~der Poel}\ \emph {et~al.}(2012)\citenamefont
  {van~der Poel}, \citenamefont {Stevens}, \citenamefont {Sugiyama},\ and\
  \citenamefont {Lohse}}]{van2012flow}%
  \BibitemOpen
  \bibfield  {author} {\bibinfo {author} {\bibfnamefont {E.~P.}\ \bibnamefont
  {van~der Poel}}, \bibinfo {author} {\bibfnamefont {R.~J.}\ \bibnamefont
  {Stevens}}, \bibinfo {author} {\bibfnamefont {K.}~\bibnamefont {Sugiyama}}, \
  and\ \bibinfo {author} {\bibfnamefont {D.}~\bibnamefont {Lohse}},\ }\bibfield
   {title} {\enquote {\bibinfo {title} {Flow states in two-dimensional
  {R}ayleigh-{B}{\'e}nard convection as a function of aspect-ratio and
  {R}ayleigh number},}\ }\href {\doibase 10.1063/1.4744988} {\bibfield
  {journal} {\bibinfo  {journal} {Physics of Fluids}\ }\textbf {\bibinfo
  {volume} {24}},\ \bibinfo {pages} {085104} (\bibinfo {year}
  {2012})}\BibitemShut {NoStop}%
\bibitem [{\citenamefont {Sugiyama}\ \emph {et~al.}(2009)\citenamefont
  {Sugiyama}, \citenamefont {Calzavarini}, \citenamefont {Grossmann},\ and\
  \citenamefont {Lohse}}]{sugiyama2009flow}%
  \BibitemOpen
  \bibfield  {author} {\bibinfo {author} {\bibfnamefont {K.}~\bibnamefont
  {Sugiyama}}, \bibinfo {author} {\bibfnamefont {E.}~\bibnamefont
  {Calzavarini}}, \bibinfo {author} {\bibfnamefont {S.}~\bibnamefont
  {Grossmann}}, \ and\ \bibinfo {author} {\bibfnamefont {D.}~\bibnamefont
  {Lohse}},\ }\bibfield  {title} {\enquote {\bibinfo {title} {Flow organization
  in two-dimensional non-{O}berbeck--{B}oussinesq {R}ayleigh--{B}{\'e}nard
  convection in water},}\ }\href {\doibase 10.1017/S0022112009008027}
  {\bibfield  {journal} {\bibinfo  {journal} {Journal of Fluid Mechanics}\
  }\textbf {\bibinfo {volume} {637}},\ \bibinfo {pages} {105--135} (\bibinfo
  {year} {2009})}\BibitemShut {NoStop}%
\bibitem [{\citenamefont {Huang}\ and\ \citenamefont
  {Zhou}(2013)}]{huang2013counter}%
  \BibitemOpen
  \bibfield  {author} {\bibinfo {author} {\bibfnamefont {Y.-X.}\ \bibnamefont
  {Huang}}\ and\ \bibinfo {author} {\bibfnamefont {Q.}~\bibnamefont {Zhou}},\
  }\bibfield  {title} {\enquote {\bibinfo {title} {Counter-gradient heat
  transport in two-dimensional turbulent {R}ayleigh--{B}{\'e}nard
  convection},}\ }\href {\doibase 10.1017/jfm.2013.585} {\bibfield  {journal}
  {\bibinfo  {journal} {Journal of Fluid Mechanics}\ }\textbf {\bibinfo
  {volume} {737}},\ \bibinfo {pages} {1--12} (\bibinfo {year}
  {2013})}\BibitemShut {NoStop}%
\bibitem [{\citenamefont {Heslot}, \citenamefont {Castaing},\ and\
  \citenamefont {Libchaber}(1987)}]{heslot1987transitions}%
  \BibitemOpen
  \bibfield  {author} {\bibinfo {author} {\bibfnamefont {F.}~\bibnamefont
  {Heslot}}, \bibinfo {author} {\bibfnamefont {B.}~\bibnamefont {Castaing}}, \
  and\ \bibinfo {author} {\bibfnamefont {A.}~\bibnamefont {Libchaber}},\
  }\bibfield  {title} {\enquote {\bibinfo {title} {Transitions to turbulence in
  helium gas},}\ }\href {\doibase 10.1103/PhysRevA.36.5870} {\bibfield
  {journal} {\bibinfo  {journal} {Physical Review A}\ }\textbf {\bibinfo
  {volume} {36}},\ \bibinfo {pages} {5870} (\bibinfo {year}
  {1987})}\BibitemShut {NoStop}%
\bibitem [{\citenamefont {Castaing}\ \emph {et~al.}(1989)\citenamefont
  {Castaing}, \citenamefont {Gunaratne}, \citenamefont {Heslot}, \citenamefont
  {Kadanoff}, \citenamefont {Libchaber}, \citenamefont {Thomae}, \citenamefont
  {Wu}, \citenamefont {Zaleski},\ and\ \citenamefont
  {Zanetti}}]{castaing1989scaling}%
  \BibitemOpen
  \bibfield  {author} {\bibinfo {author} {\bibfnamefont {B.}~\bibnamefont
  {Castaing}}, \bibinfo {author} {\bibfnamefont {G.}~\bibnamefont {Gunaratne}},
  \bibinfo {author} {\bibfnamefont {F.}~\bibnamefont {Heslot}}, \bibinfo
  {author} {\bibfnamefont {L.}~\bibnamefont {Kadanoff}}, \bibinfo {author}
  {\bibfnamefont {A.}~\bibnamefont {Libchaber}}, \bibinfo {author}
  {\bibfnamefont {S.}~\bibnamefont {Thomae}}, \bibinfo {author} {\bibfnamefont
  {X.-Z.}\ \bibnamefont {Wu}}, \bibinfo {author} {\bibfnamefont
  {S.}~\bibnamefont {Zaleski}}, \ and\ \bibinfo {author} {\bibfnamefont
  {G.}~\bibnamefont {Zanetti}},\ }\bibfield  {title} {\enquote {\bibinfo
  {title} {Scaling of hard thermal turbulence in {R}ayleigh-{B}{\'e}nard
  convection},}\ }\href {\doibase 10.1017/S0022112089001643} {\bibfield
  {journal} {\bibinfo  {journal} {Journal of Fluid Mechanics}\ }\textbf
  {\bibinfo {volume} {204}},\ \bibinfo {pages} {1--30} (\bibinfo {year}
  {1989})}\BibitemShut {NoStop}%
\bibitem [{\citenamefont {Xi}, \citenamefont {Lam},\ and\ \citenamefont
  {Xia}(2004)}]{xi2004laminar}%
  \BibitemOpen
  \bibfield  {author} {\bibinfo {author} {\bibfnamefont {H.-D.}\ \bibnamefont
  {Xi}}, \bibinfo {author} {\bibfnamefont {S.}~\bibnamefont {Lam}}, \ and\
  \bibinfo {author} {\bibfnamefont {K.-Q.}\ \bibnamefont {Xia}},\ }\bibfield
  {title} {\enquote {\bibinfo {title} {From laminar plumes to organized flows:
  the onset of large-scale circulation in turbulent thermal convection},}\
  }\href {\doibase 10.1017/S0022112004008079} {\bibfield  {journal} {\bibinfo
  {journal} {Journal of Fluid Mechanics}\ }\textbf {\bibinfo {volume} {503}},\
  \bibinfo {pages} {47--56} (\bibinfo {year} {2004})}\BibitemShut {NoStop}%
\bibitem [{\citenamefont {Kolmogorov}(1962)}]{kolmogorov1962refinement}%
  \BibitemOpen
  \bibfield  {author} {\bibinfo {author} {\bibfnamefont {A.~N.}\ \bibnamefont
  {Kolmogorov}},\ }\bibfield  {title} {\enquote {\bibinfo {title} {A refinement
  of previous hypotheses concerning the local structure of turbulence in a
  viscous incompressible fluid at high {R}eynolds number},}\ }\href {\doibase
  10.1017/S0022112062000518} {\bibfield  {journal} {\bibinfo  {journal}
  {Journal of Fluid Mechanics}\ }\textbf {\bibinfo {volume} {13}},\ \bibinfo
  {pages} {82--85} (\bibinfo {year} {1962})}\BibitemShut {NoStop}%
\bibitem [{\citenamefont {Zhou}\ and\ \citenamefont
  {Jiang}(2016)}]{zhou2016kinetic}%
  \BibitemOpen
  \bibfield  {author} {\bibinfo {author} {\bibfnamefont {Q.}~\bibnamefont
  {Zhou}}\ and\ \bibinfo {author} {\bibfnamefont {L.-F.}\ \bibnamefont
  {Jiang}},\ }\bibfield  {title} {\enquote {\bibinfo {title} {Kinetic and
  thermal energy dissipation rates in two-dimensional {R}ayleigh-{T}aylor
  turbulence},}\ }\href {\doibase 10.1063/1.4946799} {\bibfield  {journal}
  {\bibinfo  {journal} {Physics of Fluids}\ }\textbf {\bibinfo {volume} {28}},\
  \bibinfo {pages} {045109} (\bibinfo {year} {2016})}\BibitemShut {NoStop}%
\end{thebibliography}%

\end{document}